\documentclass[12pt]{iopart}

\usepackage{graphicx}
\usepackage{xcolor}
\usepackage{bm}

\newcommand{\ket}[1]{\vert {#1} \rangle}

\begin{document}
	

\title{The Structure of Phase Diagrams in Matter-Radiation Systems}

\author{Eduardo Nahmad-Achar, Sergio Cordero, Ram\'on L\'opez-Pe\~na}

\address{Instituto de Ciencias Nucleares, Universidad Nacional Aut\'onoma de
 M\'exico, Apdo.\ Postal 70-543, 04510, Cd.Mx., Mexico}
\ead{nahmad@nucleares.unam.mx}

\begin{abstract}
\vspace{0.1in}

We present a study of the structure of phase diagrams for matter-radiation systems, based on the use of coherent states and the catastrophe formalism, that compares very well with the exact quantum solutions as well as providing analytical expressions. Emphasis is made on $2$- and $3$-level systems, but in general $n$-level systems in the presence of $\ell$ electromagnetic modes are described. Due to the infinite-dimensional nature of the Hilbert space, and using the results of the analyses and the behaviour of the solutions, we construct a sequence of ever-approximating reduced bases, which make possible the study of larger systems both, in the number of atoms and in the number of excitations. These studies are of importance in fundamental quantum optics, quantum information, and quantum cryptography scenarios.

\end{abstract}

\section{Introduction}

With the ability to manipulate single atoms and photons in a cavity came a renewed interest in the models that describe their behaviour. An important feature of atom-field interactions is the presence of a phase transition from a {\it normal} to a {\it collective} behaviour: effect involving all $N$ atoms in the sample, where the decay rate is proportional to $N^2$ instead of $N$ (the expected result for independent atomic emission)~\cite{hepp73}.

Except in the thermodynamic limit systems cannot be solved analytically, so good approximations through catastrophe theory become a very useful tool of study~\cite{cordero13}. In this, the use of the Glauber coherent states~\cite{glauber63}, introduced in the context of quantum electrodynamics to provide a complete description of coherence in the electromagnetic field, has proved to be of utmost importance. They constitute the backbone of quantum optics and they have been generalised to other bosonic quantum field theories. Other studies by Glauber and co-workers (cf., e.g.,~\cite{glauber84}) include systems of coupled harmonic oscillators with thermal baths at different temperatures, where the phase diagram techniques presented in this work could be useful in order to study intermediate equilibrium states.

The study of quantum phase transitions has received much interest. The order of a quantum phase transition may be determined either by following the Ehrenfest method, through the fidelity of neighbouring states, or by means of the Wehrl entropy~\cite{castanos15}. It is also of importance, amongst other fields, in quantum information processing; entanglement measures have been used as a signature to characterise different quantum phases in models such as the Lipkin-Meshkov-Glick and the Dicke models~\cite{calixto17}.

In this work, we will analyse the structure of the phase diagram of a system of atoms in the presence of a radiation field, with particular but not exclusive interest in the case of a {\it finite} number of atoms. We treat, for ease of reading and motivation, $2$- and $3$-level atoms and $1$ or $2$ modes of the electromagnetic field, but we then generalise our results to $n$-level atoms and $\ell$ modes.

An overdetermined basis of coherent states for the Hilbert space is used, which we then adapt to maintain the symmetries of the Hamiltonian of the system. With this, we calculate the minimum energy surface in the space of the matter-field coupling parameters of the system, in order to analyse the properties of the ground state.

We also discuss a method for building an ever approximating sequence of bases for the Hilbert space of the system, which makes it much more manageable and allows us to approximate the exact quantum solution as much as is desired, as well as to tackle previously intractable problems due to the large dimension of the quantum systems.

The paper is organised as follows. Section~\ref{sec2} presents the mathematical model that describes the system. Section~\ref{sec3} takes coherent states as trial states and, via a variational procedure, obtains the critical values of the field and matter parameters. This leads to a structure of the phase diagram, which we discuss, and some expectation values are calculated and compared with those of the exact quantum solution.  In section~\ref{sec4} the symmetries of the Hamiltonian are studied and {\it symmetry-adapted states}, which preserve the Hamiltonian symmetry, are introduced. It is shown that the variational results for the ground state obtained from these states constitute an excellent approximation to the quantum solution. In section~\ref{sec5} we present a generalisation to $3$-level systems, study their phase transitions, and show that one of the atomic configurations, the $\Xi$-configuration, is special in that it presents a true triple quantum phase transition, independent of the number of atoms and constituting the thermodynamic limit of all other triple points. We show the behaviour of the ground state around this triple point, and calculate a critical exponent for the system. Section~\ref{sec6} generalises the study to the most general case of $n$-level atomic systems in the presence of a radiation field of $\ell$ modes. We show here that an iterative procedure may be carried out in order to reduce any system to $2$-level subsystems, thus simplifying the study of its phase diagram and phase transitions. In section~\ref{sec7} we briefly describe how to construct a sequence of ever-approximating bases for the Hilbert space, which allows us to overcome the strongest limitation of all: that of the exploding dimension of the Hilbert space when the number of atoms or the number of excitations grow. We close with some remarks and conclusions.

{\it This work is dedicated to the memory of Professor Roy Glauber, for his numerous contributions to the development and promotion of quantum optics and mathematical physics.}

\section{The Model for $2$-Level Systems}
\label{sec2}

\begin{figure}
	\begin{center}
		\includegraphics[width=5in]{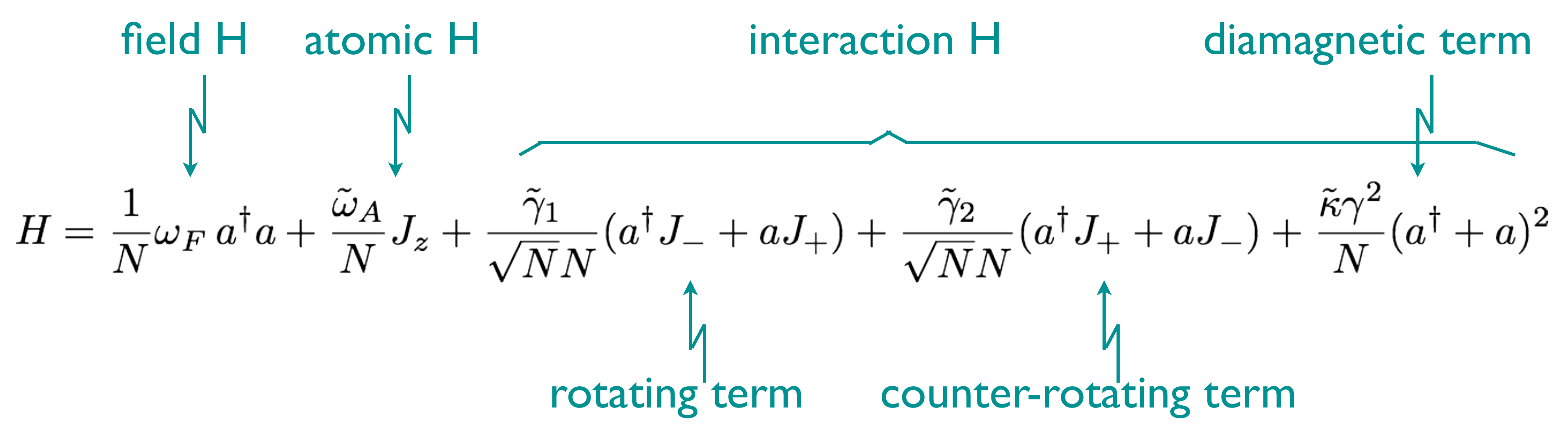}
	\end{center}
	\caption{Full Hamiltonian per particle for the interaction of a $2$-level atomic cloud with a one-mode radiation field. See text for details of each quantity. We have taken $\hbar=1$.}
	\label{FullHamiltonian}
\end{figure}

A many-body system (e.g. a cold $2$-level atomic cloud) interacting with a $1$-mode radiation field inside an optical cavity in the dipolar approximation is described by the Hamiltonian $H$ shown in Figure~\ref{FullHamiltonian}, where we have pointed out the contributions of the field, the atomic sector, and the interaction between the two.
%
Here, $\omega_F$ represents the frequency of the electromagnetic mode, $a$ and $a^\dagger$ are the annihilation and creation operators for the field, $\hbar \tilde{\omega}_A$ is the energy difference between the atomic levels, $J_z$ the population difference between these levels, $J_+$ and $J_-$ the raising and lowering atomic level operators, which satisfy the angular momentum algebra, $N$ the number of $2$-level systems (atoms or artificial atoms or spin systems), and $\gamma,\ \tilde{\gamma}_1,\ \tilde{\gamma}_2$ are coupling constants between the matter and the field. The term containing $\tilde{\kappa}$ is the so-called {\it diamagnetic term}, arising from the square of the electromagnetic vector potential $A$ upon quantisation. 

We can make the diamagnetic term vanish via the (unitary) gauge transformation~\cite{power59,siva01} $U = \exp[i\frac{e}{\hbar c} \sum_{s=1}^{N} r_s \cdot A]$, and when $\tilde{\gamma}_1 = \tilde{\gamma}_2$ we have the known Dicke model~\cite{dicke54}. Furthermore, in the rotating wave approximation (RWA), where the counter-rotating term (which does not preserve the total number of excitations) is neglected, this Hamiltonian yields the Tavis-Cummings model~\cite{tavis69}.

In this work we use the Dicke and Tavis-Cummings models, and their generalisations to accommodate any number of atomic levels and any number of field modes. It serves to work with dimensionless quantities, thus we set $\hbar=1$ and define
$$
\omega_A = \frac{\tilde{\omega}_A}{\omega_F}, \quad \gamma = \frac{\tilde{\gamma}}{\omega_F}, \quad \omega_F = 1,
$$
which allows us to measure all frequencies in units of the field frequency. We also consider indistinguishable particles, so that
$$
J = \frac{N}{2}
$$
where $J$ is the total angular momentum operator. Distinguishable particles have been considered for $2$- and $3$-level systems, using different representations for $SU(2)$ and $SU(3)$ respectively, leading to different cooperation numbers~\cite{quezada17,quezada18}.

The Dicke Hamiltonian then takes the form
\begin{equation}
H = \frac{1}{N} a^\dagger a + \frac{\omega_A}{N} J_z + \frac{\gamma}{N \sqrt{N}}(a^\dagger + a) (J_{+} + J_{-}) \nonumber
\end{equation} 
with now
$$
\omega_A,\, \gamma,\ J_{+},\, J_{-},\, J_z,\, J_x,\, J_y:\ \ \hbox{all dimensionless}
$$
Expressions for the dimensional and dimensionless quantities are given in Table~\ref{DimensionlessTable}. This Hamiltonian is invariant under the canonical transformation $\gamma \to -\gamma$ and $a \to -a$, so all our results for expectation values and fluctuations will present this symmetry.

This system is not solvable analytically except in very special cases, so one may solve via numerical diagonalisation for specific scenarios.

\begin{table}
	\caption{Relation between dimensional and dimensionless quantities in the Hamiltonian. Here, $d$ is the atomic dipole moment, $e$ is the electron charge, $m$ the atomic mass, and $\rho$ the matter density within the quantisation volume.}
	\label{DimensionlessTable}
\begin{center}
\begin{tabular*}{0.5\linewidth}{c | c}
dimensional & dimensionless \\[1mm]
\hline
& \\[-3mm]
	$\tilde\gamma = \tilde{\omega}_A\, d \sqrt{\frac{2\pi\rho}{\hbar\omega_F}}\ \ \ \hbox{[freq]}$ &
	$\gamma = \omega_A\, d\, \sqrt{\frac{2\pi\rho}{\hbar\omega_F}}$ \\[2mm]
		$\tilde\kappa\tilde{\gamma}^2 = \frac{e^2}{2m}\frac{2\pi\rho}{\omega_F\,N}\ \ \ \hbox{[freq]}$ &
		$\kappa\gamma^2 = \frac{e^2 \pi\rho}{m \omega_F^2\,N}$ \\[2mm]
		$\tilde\kappa = \frac{e^2 \hbar}{2m d^2 \tilde{\omega}_A^2\,N}\ \ \ \hbox{1/[freq]}$ &
		$\kappa = \frac{e^2 \hbar}{2m d^2 \omega_A \omega_F\,N}$
\end{tabular*}
\end{center}
\end{table}

\section{Coherent States as Trial States}
\label{sec3}
	
Another approach is to take as a test state a direct product of coherent Heisenberg-Weyl $HW(1)$-states $\ket{\alpha}$ for the electromagnetic field, and coherent $SU(2)$-states $\ket{\zeta}$ for the atomic field
\begin{equation}	
	      \ket{\alpha}
	      \otimes\ket{\zeta} = \frac{e^{
	      -\left|\alpha\right|^{2}/2}}{
	      \left(1+\left|\zeta\right|^{2}\right)^{j}}
	      \sum_{\nu=0}^{\infty}\,\sum_{m=-j}^{+j} \left\{
	      \frac{\alpha^{\nu}}{\sqrt{\nu!}}\,
	      {2j\choose j+m}^{1/2}
	      \zeta^{j+m}\,\,\ket{\nu}\otimes\ket{j,\,m}
	      \right\}\,.
\end{equation}
The {\it energy surface} is the expectation value of the Hamiltonian in this state, ${\cal E}(\alpha,\,\zeta) \equiv 
  \langle\alpha\vert\otimes\langle\zeta\vert\,\,
  H\,\,\vert\alpha\rangle\otimes\vert\zeta\rangle$
and is given by~\cite{castanos09}
\begin{equation}
      {\cal E}(\alpha,\,\zeta) = \frac{1}{2}\left(p^{2}+q^{2}\right)-j \,
      \omega_{A}\,\cos\theta+2\sqrt{j}\gamma\,q\,\sin\theta\,
      \cos\phi\ ,
   \end{equation}
where we have defined
  \begin{equation}
      \alpha = \frac{1}{\sqrt{2}}\left(q+i\,p\right)\ , \qquad
      \zeta = \tan\left(\frac{\theta}{2}\right)
      \,\exp\left(i\,\phi\right)\ .
   \end{equation}
\noindent
Here $(q,\,p)$ correspond to the expectation values of the radiation field quadratures, and $(\theta,\,\phi)$ determine a point on the Bloch sphere.
	
Critical points which minimise the energy surface are obtained via a variational procedure on these variables. These are found to be
\begin{equation*}
\fl	\left\{
\begin{array}{ll}
	\theta_{c}=0,\pi,\ q_{c}=0,\ p_{c}=0, & \vert\gamma\vert<\gamma_{c}\ ;\\

	\theta_{c}=\arccos(\gamma_c/\gamma)^{2},\ 
		q_{c}=-2\,\sqrt{j}\,\gamma\,
		\sqrt{1-(\gamma_c/\gamma)^{4}}\cos{\phi_c},\ p_{c}=0,\ \phi_{c}=0,\,\pi,
		& \vert\gamma\vert>\gamma_{c}\ .
\end{array}
	\right.
\end{equation*}

The critical points of $\mathcal{E}$ also determine $3$ regions, viz.,
\begin{equation}
\fl	\left\{
	\begin{array}{lll}
	\theta_{c}=0\,,&E_{0}=-\frac{N\,\omega_{A}}{2}\,,&\lambda_{c}=0 \nonumber\\
	\theta_{c}=\pi\,,&E_{0}=\phantom{-}\frac{N\,\omega_{A}}{2}\,,&\lambda_{c}=N \nonumber\\
	\theta_{c}=\arccos\left(\frac{\omega_{A}}{\gamma^{2}}\right)\,,
	&E_{0}=-\frac{N(\omega_{A}^{2}+\gamma^{4})}{4\,\gamma^{2}}\,,&
	\lambda_{c}=\frac{N(-\omega_{A}\,\left(\omega_{A}+2\right)+\gamma^{4}+2\gamma^2)}{4\, \gamma^{2}}
	\end{array}
	\right.
\end{equation}
for $\omega_{A}>4\gamma^2$,  $\omega_{A}<-4\gamma^2$, and  $\left|\omega_{A}\right|<4\gamma^2$ respectively

Here, $\lambda_c = \langle a^\dagger a + J_z \rangle_c + j = \langle\Lambda\rangle_c$ with $\Lambda = \sqrt{J^2 + 1/4} - 1/2 + J_z + a^\dagger a$ the total excitation number operator (a constant of motion in the RWA, and of conserved parity in the full Dicke model).

\begin{figure}
	\begin{center}
			\includegraphics[width=3in]{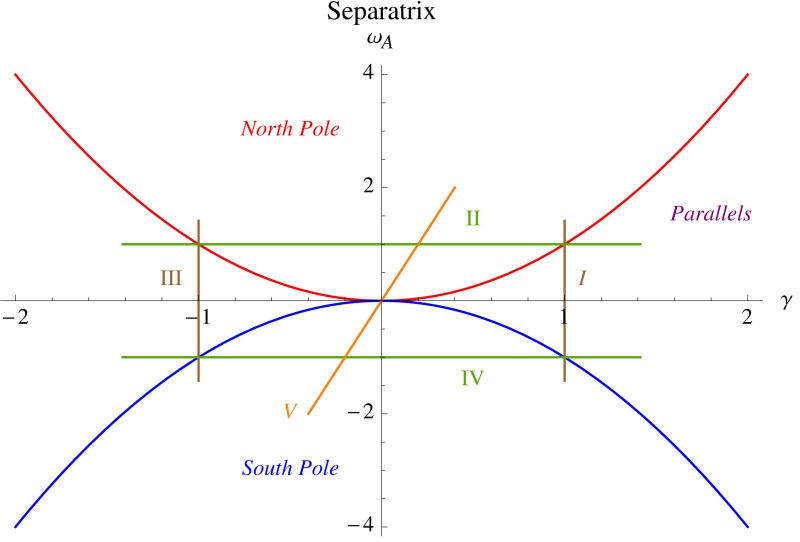}
			\caption{Separatrix for the Dicke model, showing the $3$ regions in parameter space ($\gamma,\,\omega_A$) corresponding to $\theta_c=0$ ({\it North Pole}), $\theta_c=\pi$ ({\it South Pole}), both normal regions, and the collective region $\theta_{c}=\arccos\left({\omega_{A}}/{\gamma^{2}}\right)$. Different paths crossing the separatrix are shown; all crossings are second order transitions except for the crossing of path $V$ at the origin, which is of first order.}
			\label{SeparatrixDicke}
	\end{center}
\end{figure}

The $3$ regions define a {\it separatrix}~\cite{castanos09,nahmad-achar13}, where the Hessian of $\mathcal{E}$ is singular, given by $\omega_A = \pm 4 \gamma_c^2$. This is shown in Figure~\ref{SeparatrixDicke}. The parameter space is ($\gamma,\,\omega_A$), and different paths crossing the separatrix are shown. Crossing the separatrix along paths $I,\ II,\ III$, and $IV$ leads to second-order phase transitions; crossing it along path $V$ to first order transitions. The figure is for the Tavis-Cummings model; in the Dicke model the parameter space is rescaled by a factor of $1/2$, with the same results.
%
	
\begin{figure}
	\begin{center}
\scalebox{0.40}{\includegraphics{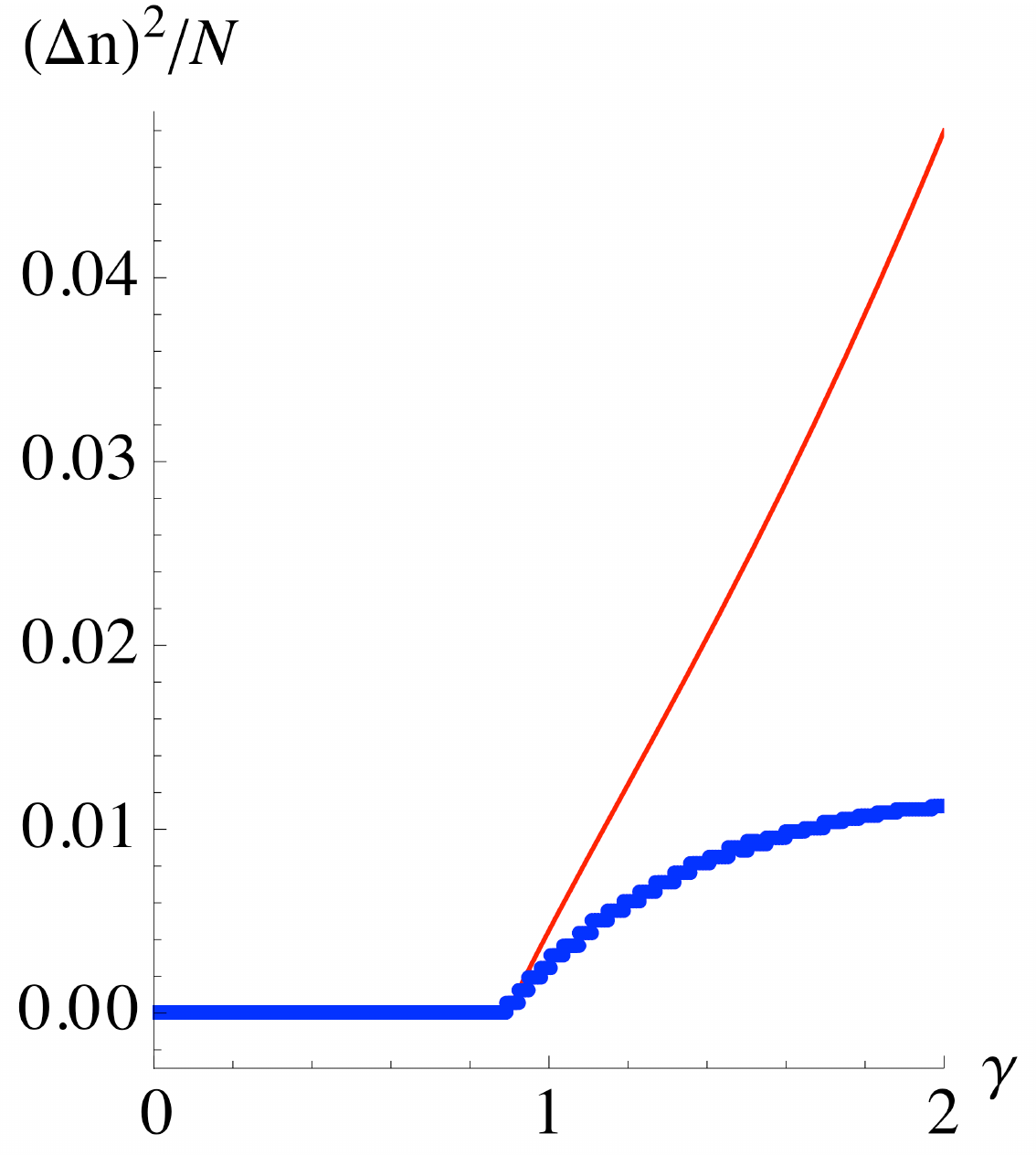}}
	\caption{The fluctuation in the number of photons plotted against the coupling parameter $\gamma$. The exact quantum solution (lower, blue curve) levels off at around $0.01$; the coherent state solution (upper, red curve) grows unboundedly.}
	\label{nf_TCM}
	\end{center}
\end{figure}

In general, coherent variational states approximate well the properties of the ground state of the quantum solution. But some properties are not well pictured; an example is shown in Figure~\ref{nf_TCM} where the fluctuation in the number of photons is plotted against the coupling parameter $\gamma$. While the exact quantum solution (lower, blue curve in the figure) levels off at around $0.01$, the coherent state solution (upper, red curve in the figure) grows unbounded.

Differences arise from the fact that the coherent state contains contributions from all eigenvalues 
\begin{equation}
	\lambda = \nu + m + j
\end{equation}
of the {\it excitations number operator} $\Lambda$, and therefore does not reflect the symmetry of the Hamiltonian leading to the constant of motion.

\section{Symmetry-Adapted States (SAS)}
\label{sec4}

Considering the unitary transformations
\begin{equation}
		U(\chi) = \exp\left(i\,\chi\,\Lambda\right)\,,\quad
		\hbox{with}\quad \Lambda = a^{\dagger}a+J_{z}
		+\frac{N}{2}\,I\,,
\end{equation}
and using the relations
\[
	U(\chi)\,a\,U^{\dagger}(\chi)
	=e^{-i\,\chi}\,a\ ,\quad
	U(\chi)\,J_{+}\,U^{\dagger}(\chi)
	=e^{-i\,\chi}\,J_{+}\ ,
\]
we find
\begin{eqnarray}
	U(\chi)\,H\,U^{\dagger}(\chi) &=& a^{\dagger}a
    +\omega_{A}J_{z}+\frac{\gamma}{\sqrt{N}}
    \left(a^{\dagger}\,J_{-}
    +a\,J_{+}\right) \nonumber\\
    &&\phantom{Un espacio}+\frac{\gamma}{\sqrt{N}}
    \left(e^{-2i\chi}\,a^{\dagger}\,J_{+}+
    e^{2i\chi}\,a\,J_{-}\right)\,,
\end{eqnarray}
so that we have a {\it symmetry transformation} for
$\chi = 0,\,\pi$, i.e., H is invariant under the group ${\cal C}_{2}=\left\{I, e^{i\,\pi\Lambda}\right\}$.

This parity symmetry
$$
	[e^{i \pi \Lambda},\, H] = 0
$$
allows for the classification of the eigenstates in terms of the parity of the eigenvalues of $\Lambda$, $\lambda = \nu + m + j$, \ldots$\!$but coherent states {\it do not} have this symmetry. We may recover the symmetries of the Hamiltonian by projecting with
\begin{equation}
			P_{\pm}=\frac{1}{2}\left(I \pm
	    e^{i\,\pi\,\Lambda}\right)
\end{equation}
thereby obtaining the so-called {\it symmetry-adapted states} (SAS)~\cite{castanos11}
	\begin{eqnarray}
		\vert\alpha,\,\zeta\rangle_{\pm}&=&
		{\cal N}_{\pm}\,P_{\pm}\,
		\vert\alpha,\,\zeta\rangle \nonumber\\
		&=&{\cal N}_{\pm}\,\left[ \ 
		\vert\alpha\rangle\otimes\vert\zeta\rangle\pm\
		\vert-\alpha\rangle\otimes\vert-\zeta\rangle\ \right]\
	\end{eqnarray} 
where ${\cal N}_{\pm}$ denotes the normalisation factor.

The expectation value of $H$ takes the form
	\begin{eqnarray}
		&&\langle H\rangle_\pm=\pm\frac{1}{2}\left(p^2+q^2\right)
		\left\{1-\frac{2}{1\pm e^{\pm(p^2+q^2)}(\cos\theta)^{\mp
		N}}\right\} \nonumber\\
		&&\quad-\frac{N}{2}\,\omega_{A}
 		\left\{(\cos \theta)^{\pm 1} 
 		\pm\frac{\tan^2\theta\,\cos\theta}{1\pm e^{\pm(p^2+q^2)} 
		(\cos\theta)^{\mp N} }\right\} \nonumber\\
		&&\quad+\sqrt{2\,N}\,\gamma\left\{\frac{\pm p\,\tan\theta\, 
		\sin\phi+q\,e^{p^2+q^2}
		\sin\theta\,\cos\phi\,(\cos\theta)^{-N}}{
		e^{p^2+q^2}(\cos\theta)^{-N}\pm 1}\right\}
\end{eqnarray}	
and is amiable to analytical calculations.

\begin{figure}
	\begin{center}
			\scalebox{0.40}{\includegraphics{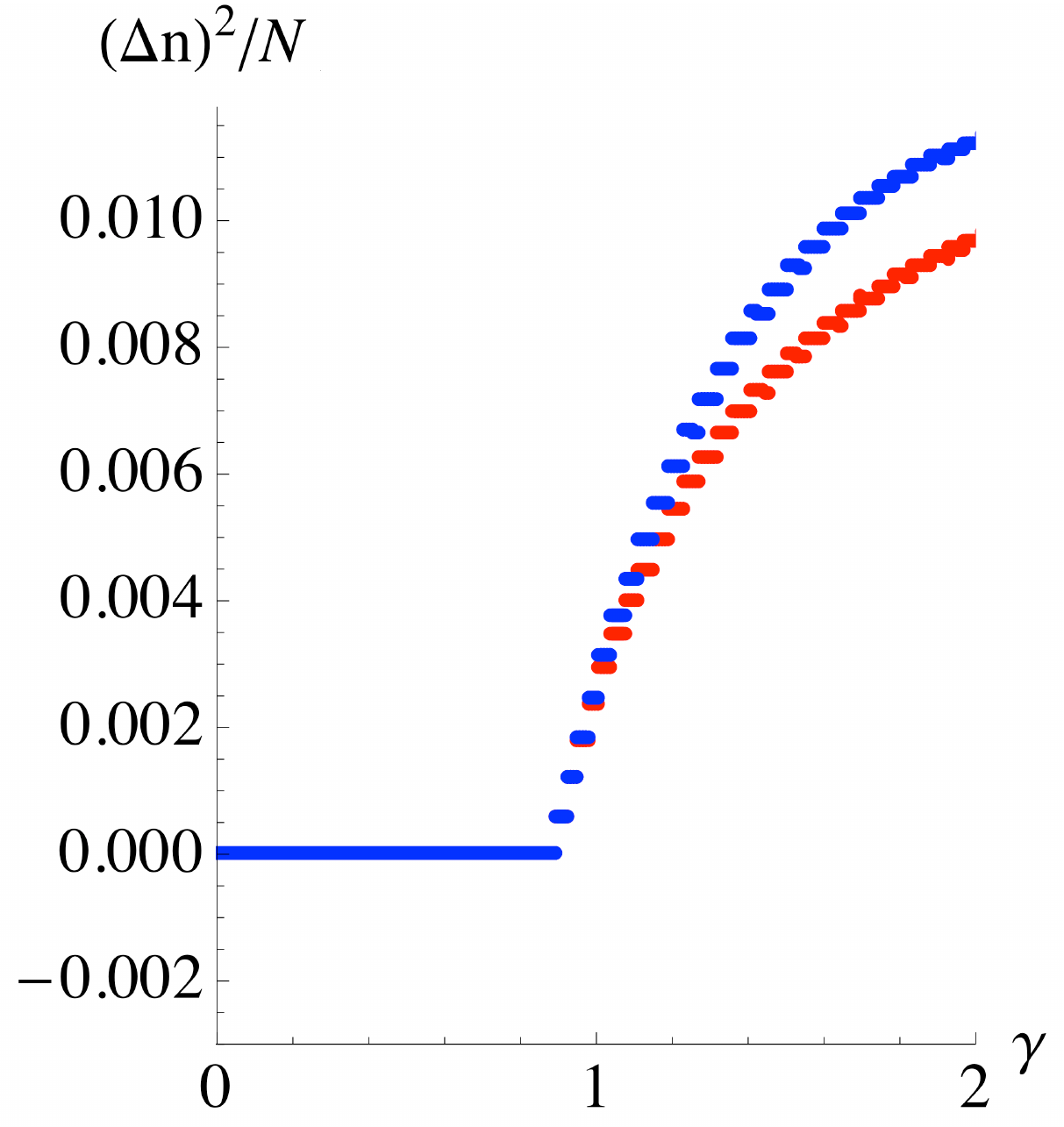}}
		\caption{The fluctuation in the number of photons, plotted against the coupling parameter $\gamma$, obtained from the symmetry-adapted ground state solution (lower, red curve), compared to that of the exact quantum solution (upper, blue curve). Note the difference in scale with respect to the previous figure where coherent states were used.}
		\label{nf_proj_TCM}
	\end{center}
\end{figure}

Working variationally with these states yields a much better approximation to the exact quantum solution in all the expectation values of field and matter operators, except in a very small vicinity of the separatrix~\cite{castanos11}; here we just show that the approximation to the photon number fluctuation is restored. In order to compare with the results in Fig.~\ref{nf_TCM}, we show in Fig.~\ref{nf_proj_TCM} the photon number fluctuation for the Tavis-Cummings model; note the scale difference in these figures. We have used, for both of them, $N=20,\ \Delta=0.2$, where the detuning is defined as $\Delta=\omega_F - \omega_A = 1 - \omega_A$.)

That restoring the Hamiltonian symmetries yields a variational basis with which one may much better resemble the exact quantum states may be verified by using the fidelity between the projected state and exact quantum ground state. This is a measure of how similar two states are, and it is given by
	\begin{equation}
		F\left(\varrho_{1},\,\varrho_{2}\right)=
		\,\hbox{tr}\left(\sqrt{\sqrt{\varrho_{1}}\,
		\varrho_{2}\,\sqrt{\varrho_{1}}}\right)\ ,
	\end{equation}

\begin{figure}
	\begin{center}
			\includegraphics[width=0.4\linewidth]{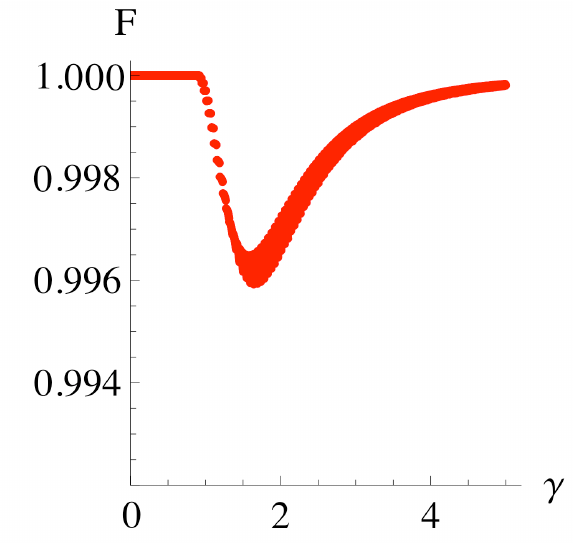}
		\caption{Fidelity between the projected ground state and the exact quantum ground state in the Tavis-Cummings model. Note that it only drops slightly at the phase transition.}
		\label{fidelity_TCM}
	\end{center}
\end{figure}

Figure~\ref{fidelity_TCM} shows, for the same $N$ and $\Delta$, the fidelity between the projected ground state and the exact quantum ground state in the Tavis-Cummings model. It is equal to $1$ in the normal region, drops to $0.996$ at the phase transition, only to recover itself at larger values of the coupling constant. The behaviour is very similar in the Dicke model.

\section{$3$-Level Systems}
\label{sec5}

\begin{figure}
	\begin{center}
			\scalebox{1.2}{\includegraphics{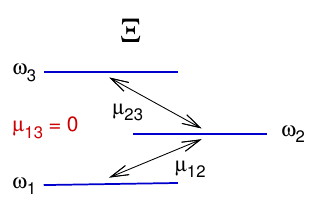}} \quad
			\scalebox{1.2}{\includegraphics{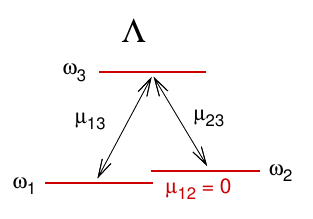}} \quad
			\scalebox{1.2}{\includegraphics{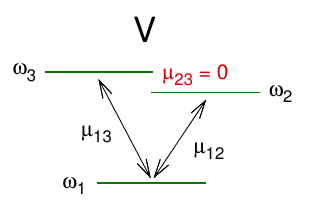}}
		\caption{Atomic configurations for $3$-level systems. $\mu_{ij}$ denotes the coupling constant between the radiation field and the transition between levels $i$ and $j$.}
		\label{f0fig}
	\end{center}
\end{figure}

$3$-level systems are richer in structure and, due to the dipolar nature of the interaction, there are $3$ atomic configurations called $\Xi$, $\Lambda$, and $V$ depending on the possible transitions, as shown in Fig.~\ref{f0fig}, where we now denote by $\mu_{ij}$ the coupling constant between the radiation field and the transition between levels $i$ and $j$, and we label the atomic levels following the convention $\omega_1 \leq \omega_2 \leq \omega_3$. 

Proposals have been made to use them as quantum memories or to manipulate quantum information, among other applications~\cite{yi03}. In cavity QED these systems have been favoured in particular because of their advantage when subjected to coherent manipulations, and schemes have been presented for various quantum gates using $3$-level atoms and trapped ions~\cite{jane02}. Furthermore, the monitoring of quantum jumps has been recently made possible using superconducting artificial $3$-level atoms~\cite{minev19}; while these continue to appear unpredictable in the long time scale, they seem to be predictable in the short time scale, and this may have applications for error correction in quantum information and computing.

For $N$ atoms of $3$ levels in the $\Xi$-configuration, interacting with a $1$-mode electromagnetic field in the dipolar and RWA approximations, the Dicke Hamiltonian generalises to
\begin{eqnarray}
H &=& \Omega\, a^\dag a + \omega_1\,A_{11} + \omega_2\,A_{22} + \omega_3\,A_{33}\nonumber \\
&-& \frac{1}{\sqrt{N}}\mu_{12}\left(a\,A_{21} + a^\dag\,A_{12} \right) 
-\frac{1}{\sqrt{N}} \mu_{23}\left(a\,A_{32} + a^\dag\,A_{23}\right)
\end{eqnarray}
where here $\Omega$ is the frequency of the field mode; $\omega_i$ is the frequency of the
$i-th$ atomic level, with $\omega_1\leq\omega_2\leq\omega_3$; ${a}^\dag,\, {a}$ are the creation and annihilation field operators; $\mu_{ij}$ are the coupling strength parameters between levels $i,\ j$; and ${A}_{ij}$ are the collective atomic transition operators, with $A_{kk}$ denoting the atomic population of level $k$.

Two operators of the form $C = \lambda\,a^\dag\,a + \lambda_1\,A_{11} + \lambda_2\,A_{22} + \lambda_3\,A_{33}$ commute with the Hamiltonian:
\begin{eqnarray*}
N &=& \sum_{i=1}^3 A_{ii} \quad \rm{total\ number\ of\ atoms}\\
M &=& a^\dag a + A_{22} + 2\, A_{33} \quad \rm{total\ number\ of\ excitations}
\end{eqnarray*}

As before, the system is not solvable analytically, so one has to solve via numerical diagonalisation for specific scenarios. A natural basis in which to diagonalise our Hamiltonian is the tensorial product of $HW(1)$ for the field sector and the  Gelfand-Tsetlin basis for the atomic sector, which in the case of totally indistinguishable atoms takes the form~\cite{cordero13b}
$$
\vert\nu;\,q,\,r\rangle = \vert\nu\rangle \otimes \vert\,q,\,r\rangle
$$
where $\nu$ labels the number of photons of the Fock state, and $r,\, q-r$ and $N-q$ are the atomic population of levels $1,\, 2,\, 3$, respectively. Since the Hamiltonian is invariant under the transformation $a\to -a, \  a^\dagger\to -a^\dagger, \ \mu_{ij}\to -\mu_{ij}$, we consider only positive values for $\mu_{ij}$.
	
The catastrophe formalism described above for $2$-level systems may be carried out here to calculate the energy of ground state as function of the coupling parameters, and the separatrices calculated via the fidelity $F$ or the fidelity susceptibility $\chi$ of neighbouring states~\cite{zanardi06,kwok08,castanos12}
\begin{eqnarray}
	&&F(\lambda,\,\lambda+\delta\lambda) = \vert\langle\psi(\lambda)\vert\psi(\lambda+\delta\lambda)\rangle\vert^2\,, \nonumber \\[0.1in]
	&&\chi = 2\,\frac{1-F(\lambda,\,\lambda+\delta\lambda)}{(\delta\lambda)^2}
\end{eqnarray}

\subsection{A Triple-Point Transition}

\begin{figure}
	\begin{center}
		\includegraphics[width=0.5\linewidth]{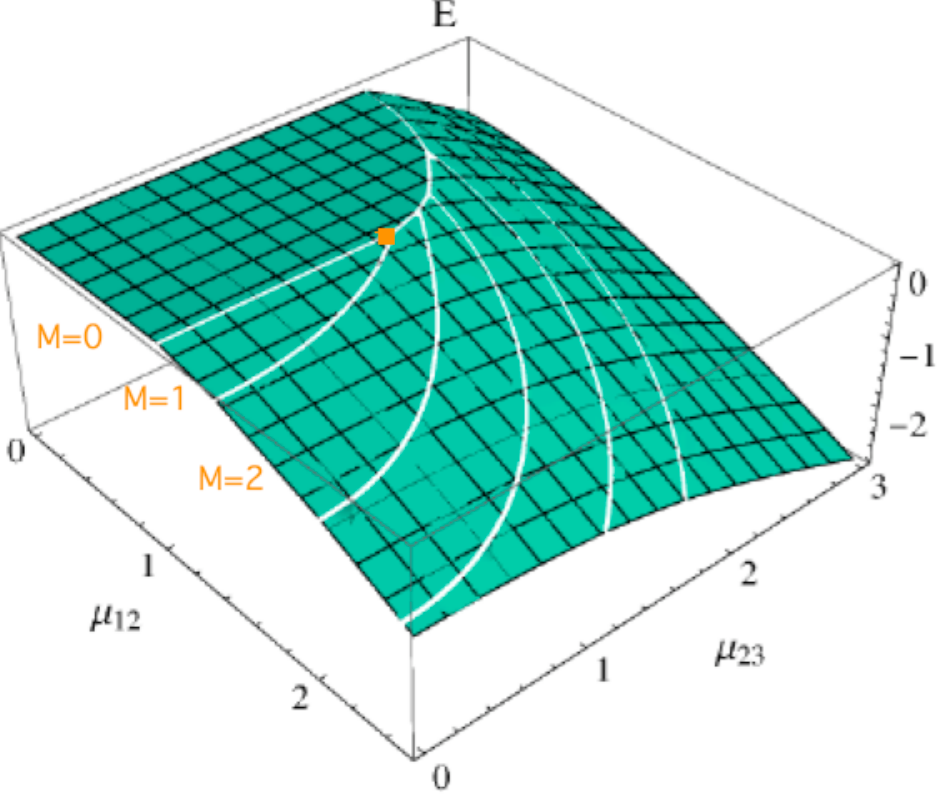}
		\caption{Surface energy for the $\Xi$-atomic configuration. White lines separate regions with different values of the total number of excitations $M$, starting at $M=0$ for small values of $\mu_{ij}$, and growing counterclockwise to $M=1,\,2\,\ldots$. Here $N=2$, $\omega_1=0$, $\omega_2=1$, $\omega_3=2$, and the system is in total resonance $\Omega=1$.}
		\label{fig3b-PTenergia2Particulas}
	\end{center}
\end{figure}

There are distinct regions for each integer value of $M$: the {\it normal} region $M=0,\ \vert\,0;\,N\,N\rangle$ where all atoms are in their ground state and there are no free photons, and the {\it collective} regions where $M \neq 0$ and which meet at separatrices shown by white lines in Fig.~\ref{fig3b-PTenergia2Particulas}. It is drawn for $2$ atoms, $N=2$; as $N$ grows, the separatrix enclosing the normal region $M=0$ remains fixed as all other separatrices slide down and to the left, asymptotically approaching the $M=0$ border. 

In the thermodynamic limit $N \to \infty$ we are left only with the first separatrix. The point at $(\mu_{12},\,\mu_{23}) = (1,\,\sqrt{2})$, where the regions $M=0,\, M=1$ and $M=2$ meet (marked in the figure, in red), remains fixed and is then a {\it true triple phase transition} independent of $N$, and the limit of all other triple points. This is a
property characteristic of the $\Xi$-configuration, and any quantum fluctuation at this triple point or in its vicinity changes drastically the composition of the ground state (cf.~\cite{nahmad-achar14}).
		
Using the full Hamiltonian (including counter-rotating terms)
\begin{eqnarray}
	H &=& \Omega\, a^\dag a + \omega_1\,A_{11} + \omega_2\,A_{22} + \omega_3\,A_{33}\nonumber \\
	&-& \frac{1}{\sqrt{N}} \left(a + a^\dag \right) \mu_{12}\left(A_{21} + A_{12} \right)
	-\frac{1}{\sqrt{N}} \left(a + a^\dag \right) \mu_{23}\left(A_{32} + A_{23}\right)
\end{eqnarray}
has the effect of shrinking the phase space by a factor of $2$:
$$
(\mu_{12},\,\mu_{23}) = (\frac{1}{2},\,\frac{1}{\sqrt{2}}\,)
$$
$M$ is no longer conserved, but its parity is, i.e., the symmetry group of the Hamiltonian is $\mathcal{C}_2 = \{\hat{1},\,\exp(i\,\pi\,M)\}$.

\subsection{Analytic Study of the Phase Diagram}

We take as a variational state a direct product of coherent $HW(1)$-states for the electromagnetic field
\begin{equation}
      \vert\alpha\}
      = \sum_{n=0}^{\infty}\,
	\frac{\alpha^{n}}{\sqrt{n!}}\,
      \vert n \rangle
\end{equation}
and $U(3)$-states for the atomic field
\begin{equation}
\vert\zeta\} := \big|[h_{1},h_{2},h_{3}]\,
    \gamma_{1},\gamma_{2},\gamma_{3}\big\}
   =e^{\gamma_{3} {A}_{21}}
\, e^{\gamma_{2} {A}_{31}} \,  e^{\gamma_{1} {A}_{32}}\, 
     \vert\,[h_{1},h_{2},h_{3}] \rangle \ ,
\label{matterstate} 
\end{equation}
where $\vert\,[h_{1},\,h_{2},\,h_{3}]\,\rangle$ represents the highest weight state of the Gelfand-Tsetlin basis in an irreducible representation of $U(3)$, and for the completely symmetric representation $h_2=h_3=0$. Minimising $\langle\alpha;\,h_1\,q_1\,r,\,\vec{\gamma}\vert\,\, H_D\,\, \vert\,\alpha;\,h_1\,q_1\,r,\,\vec{\gamma}\rangle$ with respect to $\alpha$ and $\vec\gamma$ yields the energy surface for the ground state.

\begin{figure}
	\begin{center}
		\includegraphics[width=2.5in]{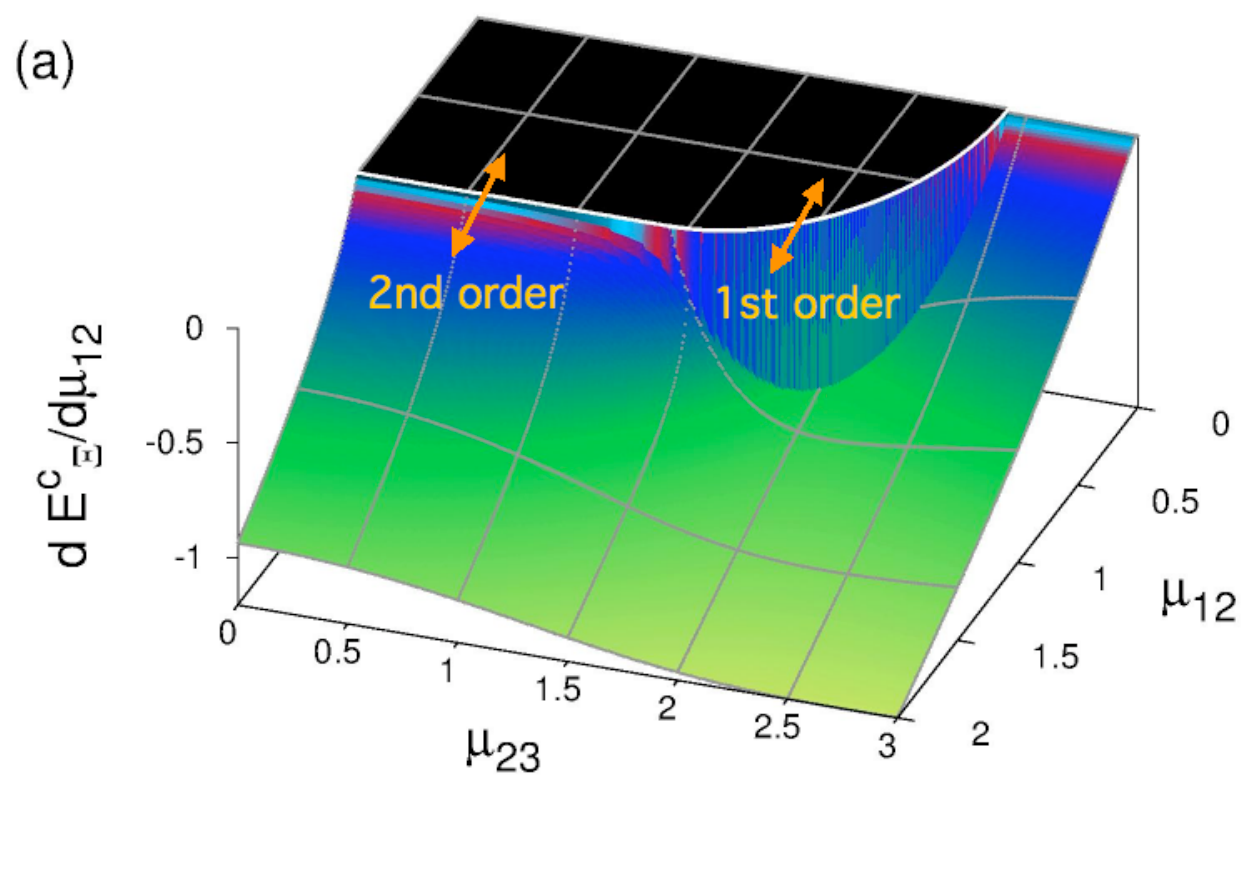} \quad
		\includegraphics[width=2.5in]{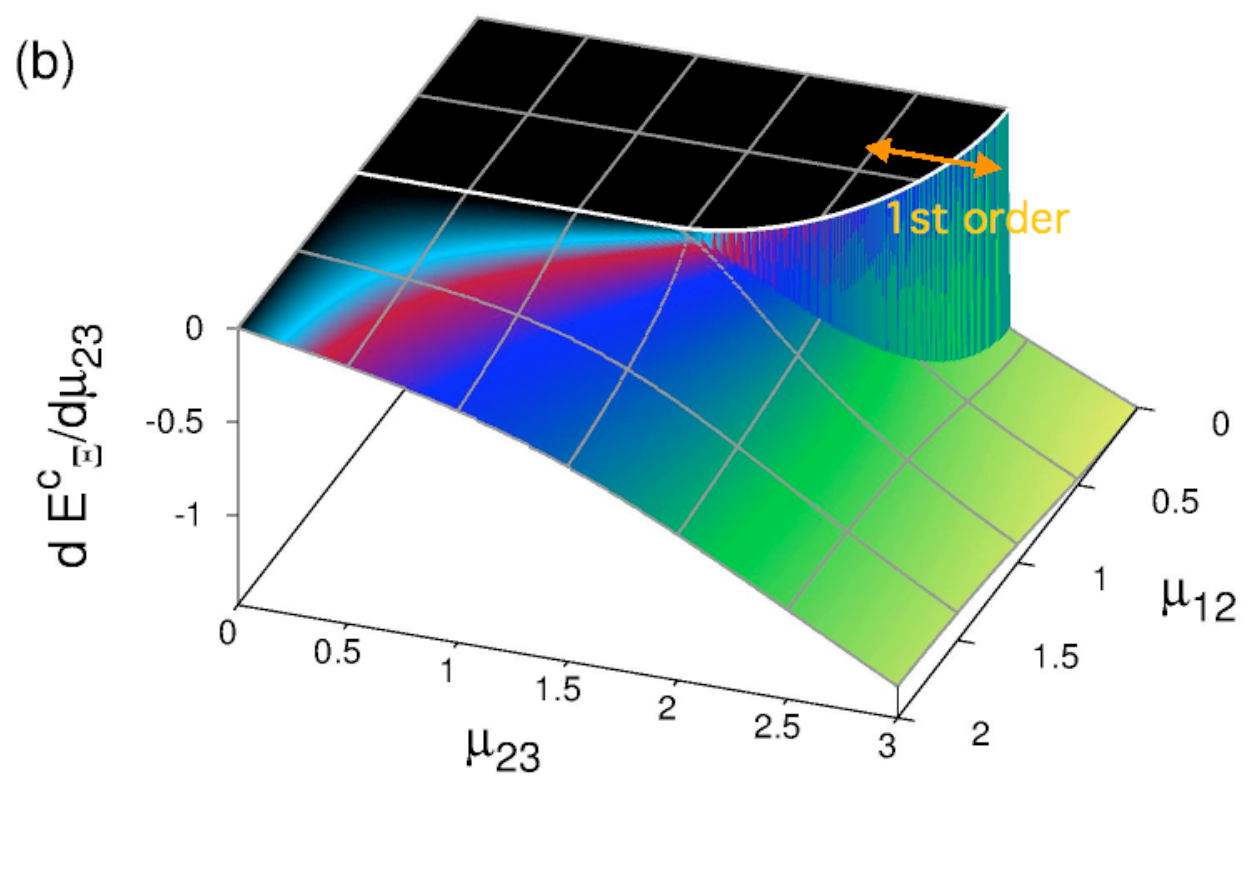}
		\caption{The derivative of the energy surface plotted as a function of the coupling parameters $(\mu_{12},\,\mu_{23})$. The order of the phase transition across the separatrix in every direction is also given.}
		\label{fig4}
	\end{center}
\end{figure}

It shows $2$ distinct regions: the normal regime $\vert\,0;\,N\,0\,N\rangle$ and the collective regime which meet at a separatrix given by
\begin{equation}
	\Omega\, \omega_{21} = \mu_{12}^2 + \left[|\mu_{23}|-\sqrt{\Omega \, \omega_{31}}\right]^2 \, \Theta\left[|\mu_{23}|-\sqrt{\Omega \, \omega_{31}}\right]
\end{equation}
with $\omega_{ij}=\omega_i - \omega_j$ and $\Theta$ the Heaviside function. This is shown as a white line in Fig.~\ref{fig4}, in which the derivative of the energy surface is plotted as a function of the coupling parameters $(\mu_{12},\,\mu_{23})$. The order of the transitions is also given, second order transitions for $\mu_{23} < \sqrt{\Omega\,\omega_3}$ and first order transitions for $\mu_{23} > \sqrt{\Omega\,\omega_3}$.

\begin{figure}
	\begin{center}
		\scalebox{0.35}{\includegraphics{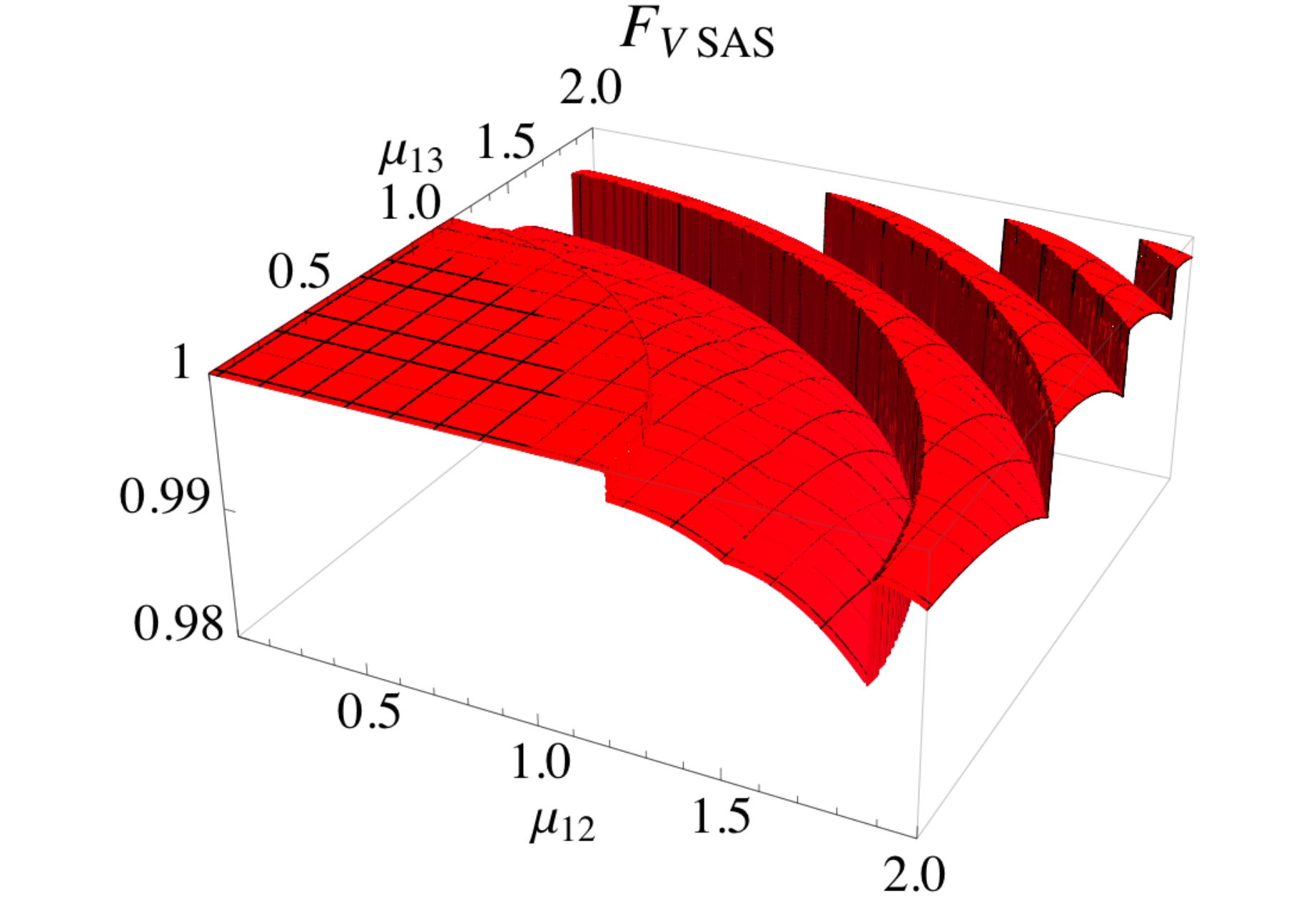}} \qquad
		\scalebox{0.35}{\includegraphics{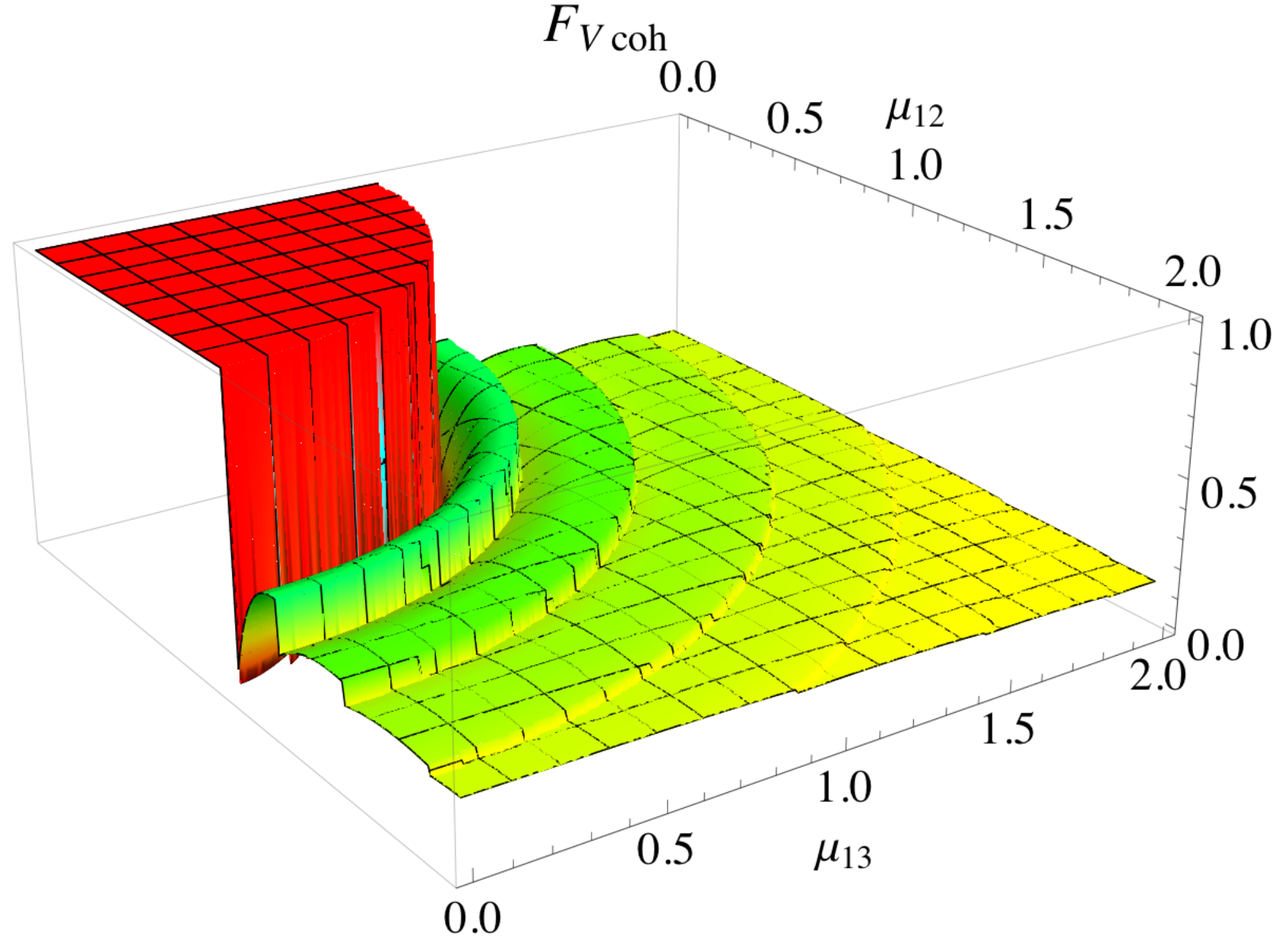}}
		\caption{Fidelity between the quantum and projected SAS ground states (left), and that between the quantum and coherent ground states (right), for $N=3$ in the $V$-configuration.}
		\label{FidelityV}
	\end{center}	
\end{figure}

Since the parity of $M$ is conserved for the quantum state, it makes sense to adapt our variational test state to a given parity
\begin{equation}
\vert\alpha,\, \vec{\gamma}\rangle_{\pm} = \left( \hat{1}\pm\exp(i\,\pi\,M)\right) \vert\alpha,\, \vec{\gamma}\rangle = \frac{1}{\sqrt{2}} \left( \vert\alpha,\, \vec{\gamma}\rangle \pm \vert -\alpha,\, \vec{\gamma}\rangle \right)
\end{equation}
where $\vert\alpha,\, \vec{\gamma}\rangle_{+}$ only contains terms with even values of $M$, and $\vert\alpha,\, \vec{\gamma}\rangle_{-}$ only contains terms with odd values of $M$. Using these, the ground and first excited SAS states give an excellent approximation to the ground and first excited quantum states. Figure~\ref{FidelityV} shows the fidelity between the quantum and projected SAS ground states (left), and that between the quantum and coherent ground states (right), for $N=3$, this time for the $V$-configuration. Notice that, except for a small vicinity of the phase transition, the SAS states do approximate very well the quantum solutions, while the coherent states do so only within the normal region but fail in the collective region.

Another good comparison between the two is given by the expectation values of the system quantities. For the number of photons, for instance, in the normal regime the coherent ground state $\vert \rm{coh}\rangle_g$ has exactly zero photons, whereas the SAS state $\vert \rm{SAS}\rangle_g$ is a superposition of states with an expectation value for $\nu$ different from zero, just as the ground state $\vert \rm{quant}\rangle_g$ is. In the limit $N\to\infty$ we have
\begin{equation}
\vert\,{}_g\langle\alpha_{coh}\,\zeta_{coh}\,\vert\,\alpha_{sas}\,\zeta_{sas}\rangle_g \,\vert^2 = \frac{1}{2}
\end{equation}
as expected: the SAS ground state has contributions only from the even-parity components of coherent ground state.

\subsection{Critical Exponents}

\begin{figure}
	\begin{center}
		\includegraphics[width=2in]{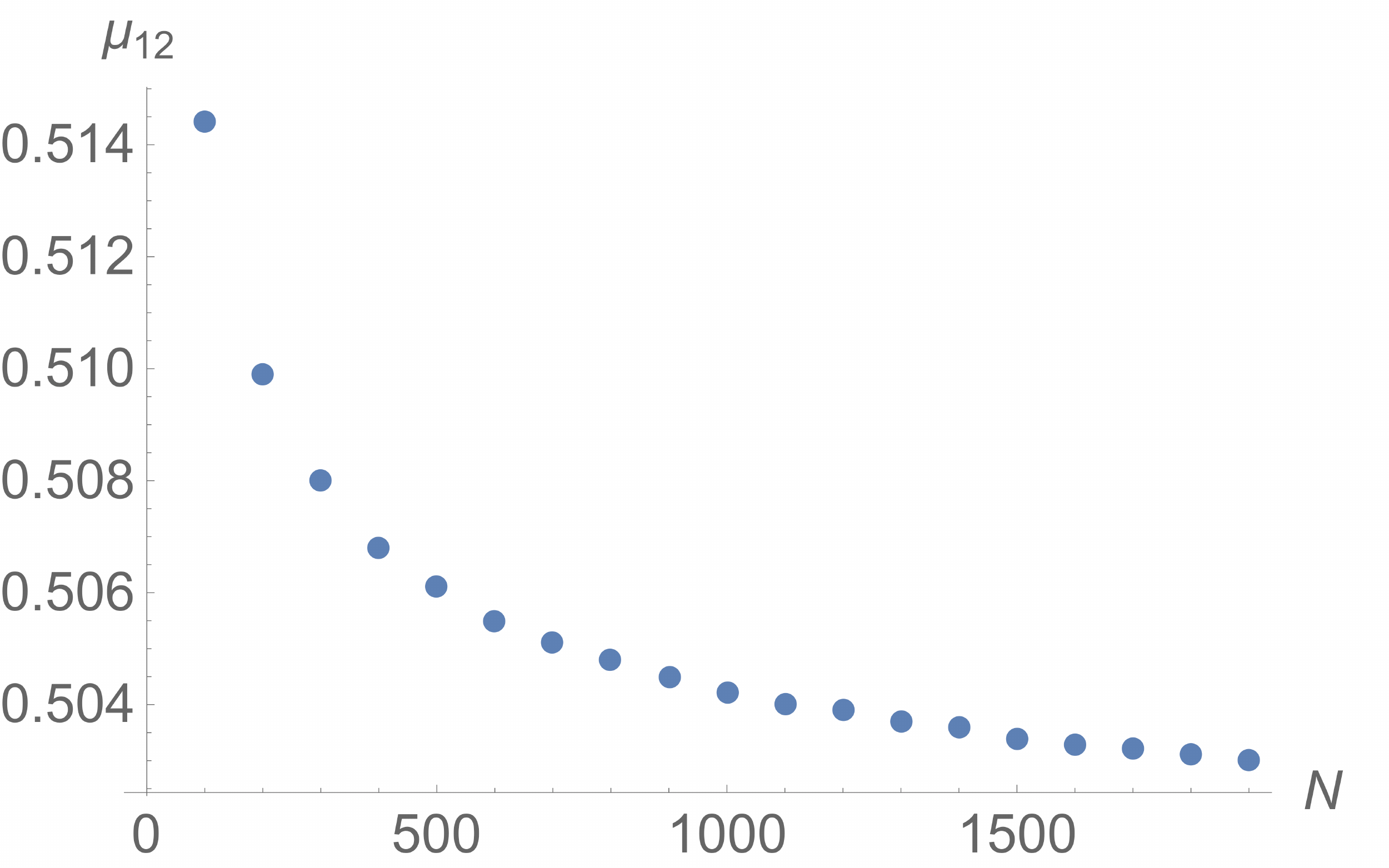} \quad
		\includegraphics[width=2in]{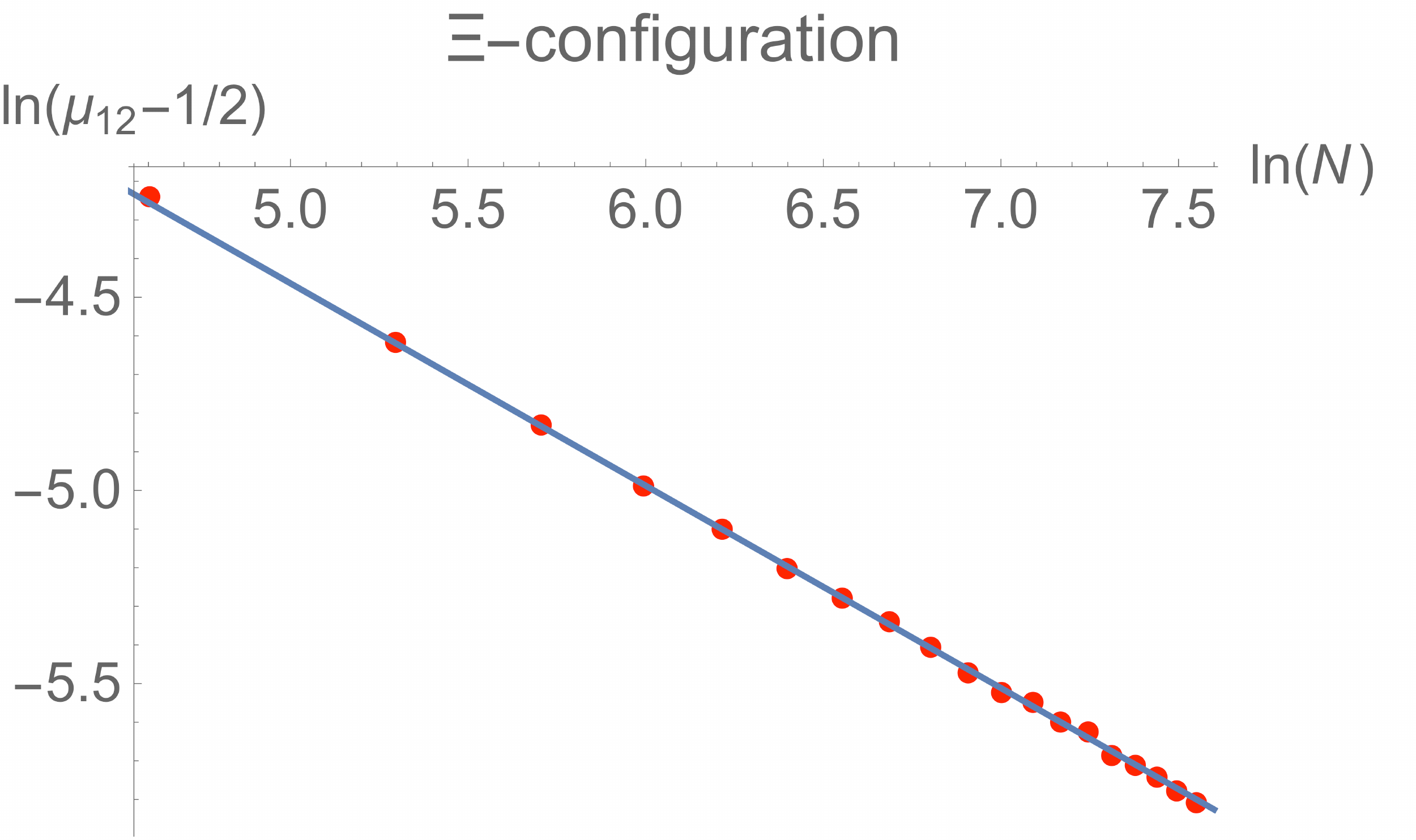}
		\caption{Critical value of the atom-field coupling parameter $\mu$ as a function of the number of atoms $N$, for the $\Xi$-configuration.}
	\label{MuvsN}
	\end{center}
\end{figure}

The singular part of many potentials in physics are homogeneous functions near second-order phase transitions
\begin{equation*}
	f(\beta r) = g(\beta)\,f(r)\, , \quad \rm{with} \quad g(\beta) = \beta^s
\end{equation*}
The behaviour of important observables of a system near phase transitions may thus be described by the system's critical exponent $s$, and these are believed to be universal with respect to physical systems. Our treatment allows us to study the critical value of the atom-field coupling parameter $\mu$ as a function of the number of atoms $N$. Although we get a very good behaviour for the SAS approximation, as shown in Fig.~\ref{MuvsN}, the exponent differs from the expected $-2/3$ in the quantum solution:
\begin{equation*}
	\ln\left(\mu_{12} - \frac{1}{2}\right) = -\frac{11}{21}\,\ln(N) + \ln(0.158)
\end{equation*}
or, equivalently,
\begin{equation}
	\mu_{12} = \frac{1}{2} + 0.158\,N{}^{-\frac{11}{21}}
\end{equation}
i.e., a critical exponent of $s_{sas}=-\frac{11}{21}$ as opposed to $s_{quant}=-\frac{2}{3}.$

\section{Generalisation to $n$ Levels and $\ell$ Modes}
\label{sec6}

In this case the Hamiltonian takes the form
$$
H = H_D + H_{int}
$$
with
\begin{equation}
H_D = \sum_{j<k}^{n} \Omega_{jk}\, a_{jk}^\dag\, a_{jk} + \sum_{j=1}^{n} \omega_j \, A_{jj}	
\end{equation}
and
\begin{equation}
H_{int} = - \frac{1}{\sqrt{N}} \sum_{j<k}^{n} \mu_{jk} \left(A_{jk}+A_{kj}\right) \left(a_{jk} + a_{jk}^\dag\right)
\end{equation}
The operators $A_{kj}$ obey a $U(n)$ algebra $\left[A_{lm},A_{kj}\right] = \delta_{mk}\,A_{lj}-\delta_{jl}\,A_{km}$, and the transition between the levels $j$ and $k$ are only promoted by mode $\Omega_{jk}$. The maximum number of dipolar interaction strengths of an $n$-level system is $\ell_{max} = n(n-1)/2- (n-2)$; of course $\ell \leq \ell_{max}$ and depends of the considered atomic configuration.

We follow, as before, a variational procedure starting from coherent states to find the energy surface, and we find the critical points with the use of the fidelity between neighbouring states to determine the separatrices~\cite{cordero15}.

\begin{figure}
	\begin{center}
		\includegraphics[width=2.5in]{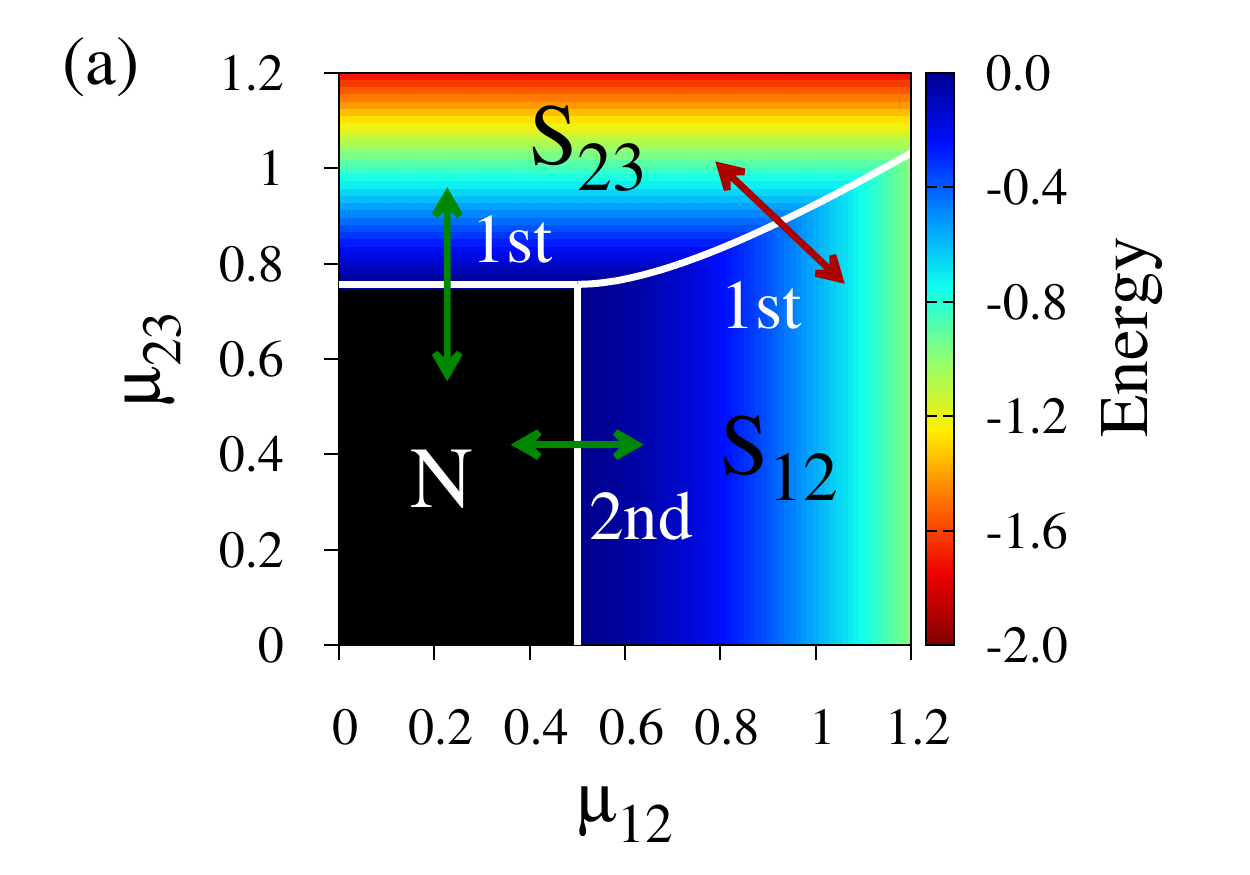} \quad
		\includegraphics[width=2.5in]{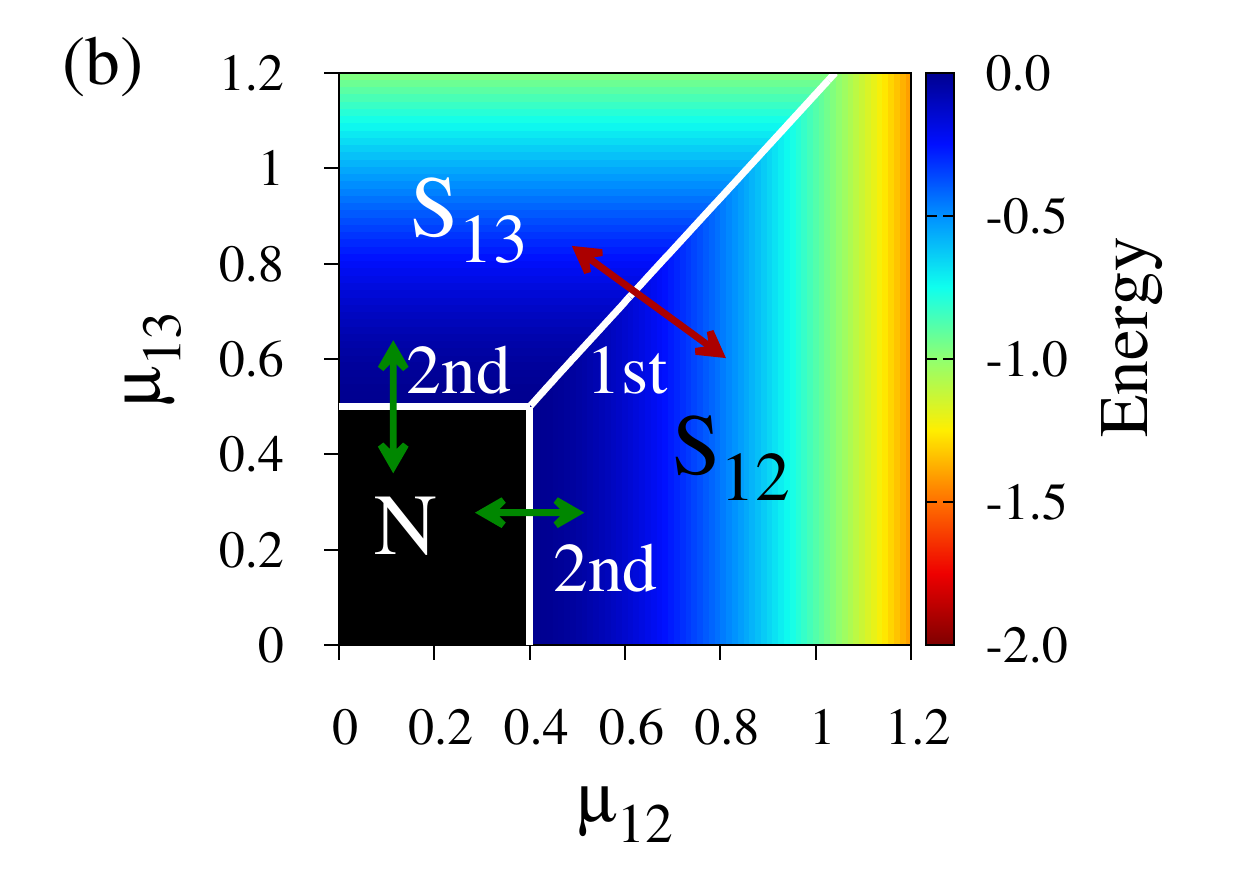} \quad
		\includegraphics[width=2.5in]{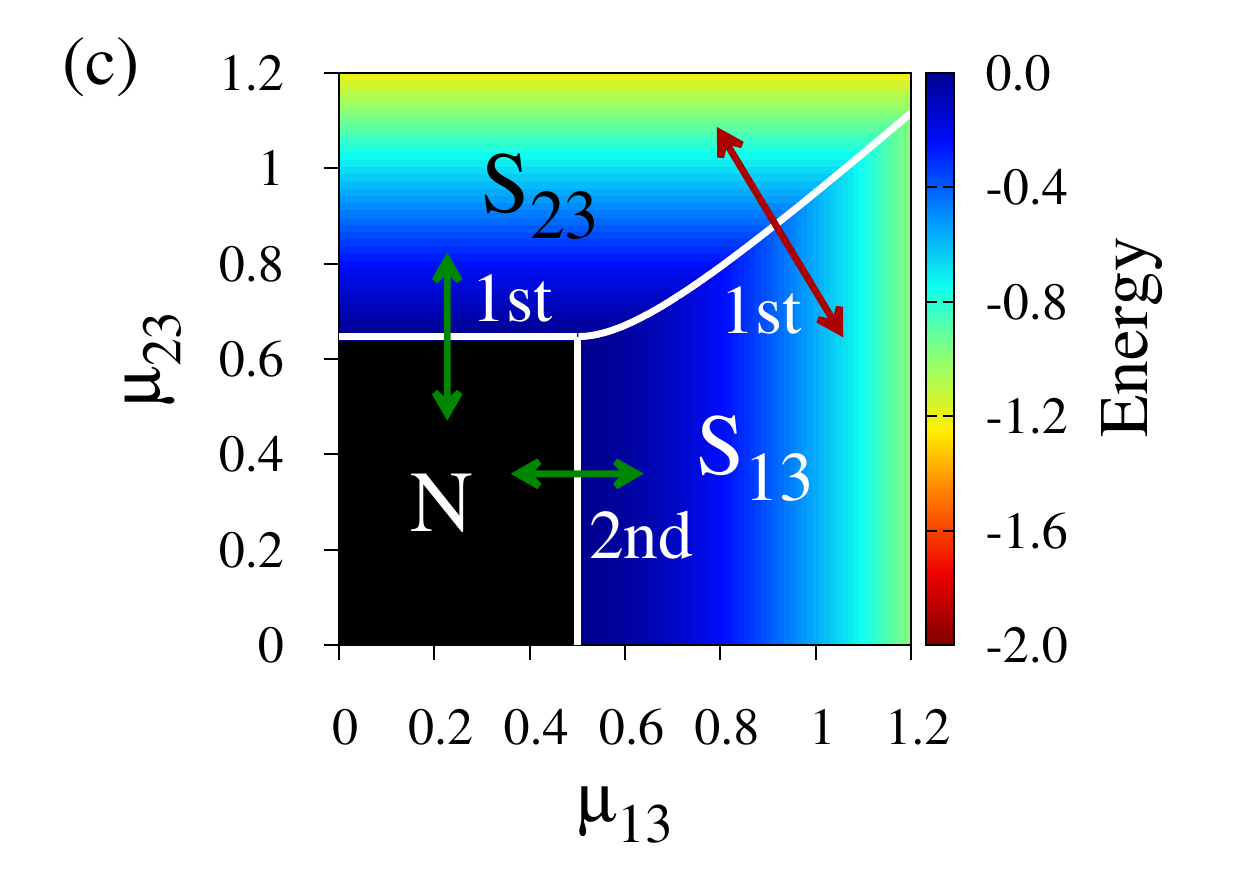}
		\caption{Phase diagram for $n=3$ levels and $\ell=2$ modes, together with the order of the transitions. N indicates the normal region (in black), and the labels $S_{ij}$ indicate that the mode $\Omega_{ij}$ dominates in these regions. a) $\Xi$-config: $\Omega_{12}=1$, $\Omega_{23}=0.5$, $\omega_1=0$, $\omega_2=1$, $\omega_3=1.3$. b) $V$-config: $\Omega_{12}=0.8$, $\Omega_{13}=1$, $\omega_1=0$, $\omega_2=0.8$, $\omega_3=1$. c) $\Lambda$-config: $\Omega_{13}=1$, $\Omega_{23}=0.8$, $\omega_1=0$, $\omega_2=0.2$, $\omega_3=1$.}
	\label{separatrizXVL}
	\end{center}
\end{figure}

Figure~\ref{separatrizXVL} shows the structure of the phase diagram for $n=3$ levels and $\ell=2$ modes, together with the order of the transitions. N indicates the normal region (in black), and the labels $S_{ij}$ indicate that the mode $\Omega_{ij}$ dominates in these regions. The parameters used are given in the figure caption.

For $n=3$ and $\ell=2$ there are $2$ parity symmetries $\Pi_j = \exp\left(i \pi K_j\right)$;
in the $\Xi$-configuration, for instance, these are
\begin{eqnarray}
K_1 &=& \nu_{12} + \nu_{23} + A_{22} + 2A_{33} \nonumber\\
K_2 &=& \nu_{23} + A_{33}
\end{eqnarray}
besides, $N = A_{11}+A_{22}+A_{33}$. It is then useful to construct symmetry-adapted states, and the Hilbert space will consist of the direct sum of $4$ sub-spaces according to the parity of each of these symmetries
\begin{equation}
	\mathcal{H} = \mathcal{H}_{ee} \oplus \mathcal{H}_{eo} \oplus \mathcal{H}_{oe} \oplus \mathcal{H}_{oo}
\end{equation}
One may calculate the energy surface in each of these sub-spaces, and then take the minimum at each point in parameter space in order to get the energy surface corresponding to the ground state. 

\begin{figure}
	\begin{center}
	\includegraphics[width=3in]{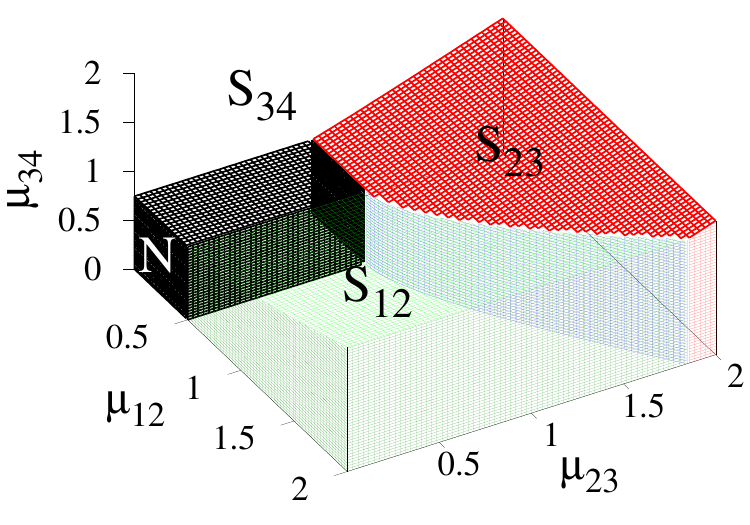}
	\caption{Phase diagram for $4$-level ladder-configuration with $\Omega_{12}=1$, $\Omega_{23}=0.7$, $\Omega_{34}=0.3$, $\omega_1=0$, $\omega_2=1$, $\omega_3=1.7$, $\omega_4=2$. The mode $\Omega_{ij}$ dominates in the region denoted by $S_{ij}$. The region $S_{34}$ lies above those coloured in the diagram. The transition N$\rightarrow S_{12}$ is of second order; all others are of first order.}
	\label{sepX4n}
\end{center}
\end{figure}

Similarly, for $n=4$ levels and $\ell=3$ modes, in the $Ladder$ configuration, the energy surface is divided into a normal region and $3$ collective regions, in each of which only a monochromatic electromagnetic field mode contributes strongly to the ground state (cf. Fig.~\ref{sepX4n}). The transition N$\rightarrow S_{12}$ is of second order; all others are of first order.

\subsection{Level Reduction}

One may allow each of the modes to interact with more than one pair of atomic levels. 
Setting $\gamma_i=\varrho_i \exp\{i\phi\}$ in equation~(\ref{matterstate}), and carrying out the variational procedure with respect to the field variables $p$ and $q$, and the matter variables $\rho_i$ ($i=1,\,2,\,3,\,4$), the critical values at $\varrho_2^c=\varrho_3^c=\varrho_4^c=0$ give the vacuum state for the field contribution and all atoms in their lower state: $\vert 0\rangle_F \otimes \vert N,0,0,0\rangle_M$. But when at least one critical value $\rho_i$ is non-zero the system may be reduced to subsystems with one number of levels less~\cite{cordero15}, from $n$ to $n-1$, essentially because we find critical points at $\infty$. Following the process iteratively, we may arrive at a collection of sub-systems of the Dicke model with one radiation mode, which in the variational method can be solved.

\begin{figure}
	\begin{center}
		\includegraphics[width=0.65\linewidth]{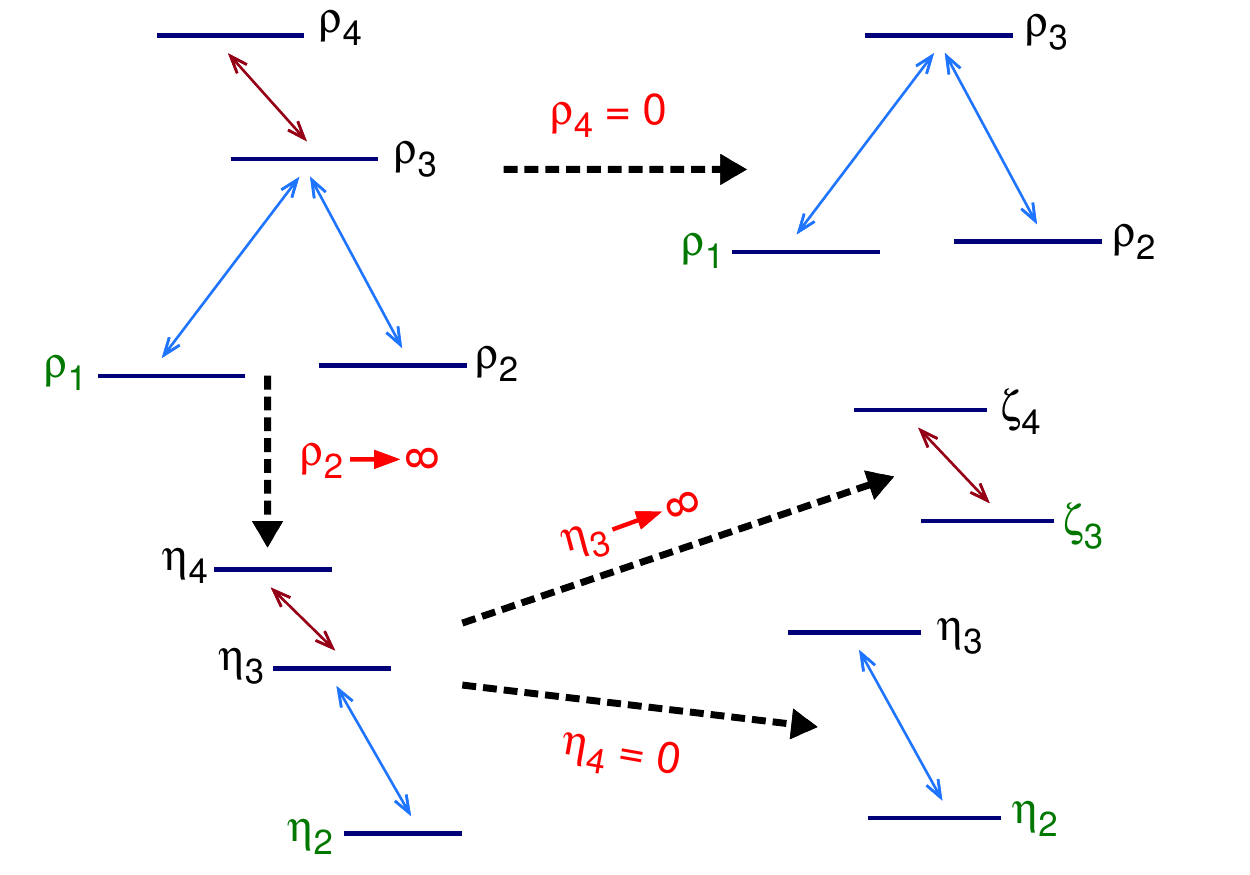}
		\caption{Reduction of the $4$-level $\lambda$-configuration to $3$- and $2$-level configurations in the collective regime. Two radiation modes are considered, one acting between levels $3 \rightleftharpoons 4$ and the other between the levels $1 \rightleftharpoons 3$ and $2 \rightleftharpoons 3$, shown in different colours.}
		\label{lambda_reduction}
	\end{center}
\end{figure}

\begin{figure}
	\begin{center}
		\includegraphics[width=0.65\linewidth]{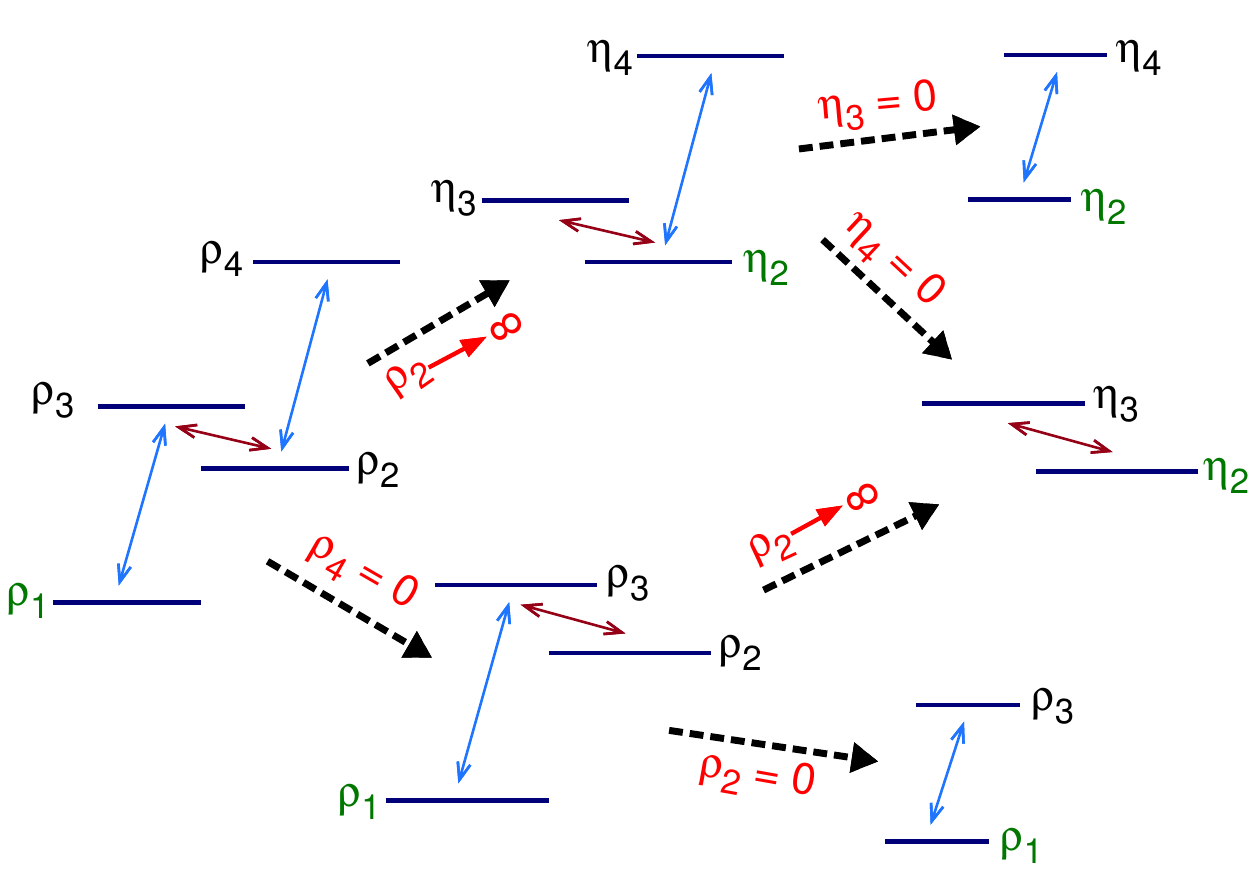}
		\caption{Reduction of the $4$-level {\tt N}-configuration to $2$-level configurations in the collective regime. Two radiation modes are considered, one acting between levels $2 \rightleftharpoons 3$ and the other between the levels $1 \rightleftharpoons 3$ and $2 \rightleftharpoons 4$, shown in different colours.}
		\label{N_reduction}
	\end{center}
\end{figure}

We have shown schematically in Figures~\ref{lambda_reduction} and~\ref{N_reduction} the reduction paths of the $4$-level configurations $\lambda$ and {\tt N}. In the case of $\lambda$, with two radiation modes (one acting between levels $3 \rightleftharpoons 4$ and the other between the levels $1 \rightleftharpoons 3$ and $2 \rightleftharpoons 3$), one may set $\rho_1=1$ for state normalisation and when $\rho_4=0$ we get the $3$-level $\Lambda$ configuration, which we can study as in previous sections. When $\rho_2\to\infty$ we obtain a $3$-level $\Xi$ configuration in new variables (denoted by $\eta$ in the figure), which itself reduces to $2$-level Dicke systems according to the critical values of $\eta_3$ and $\eta_4$.

The way in which the {\tt N} configuration splits is richer. We take here two radiation modes as well, one acting between levels $2 \rightleftharpoons 3$ and the other between the levels $1 \rightleftharpoons 2$ and $3 \rightleftharpoons 4$. Here, once again set $\rho_1=1$ and when $\rho_2\to\infty$ we get a $3$-level $V$ configuration in new variables $\eta$; this yields two $2$-level subsystems when $\eta_3=0$ and when $\eta_4=0$. On the other hand, when $\rho_4=0$ the {\tt N}-configuration reduces to a $3$-level $\Lambda$ configuration, which again reduces by iteration to two $2$-level subsystems when $\rho_2=0$ and when $\rho_2\to\infty$ (see the figure).

\begin{figure}
	\begin{center}
		\includegraphics[width=0.6\linewidth]{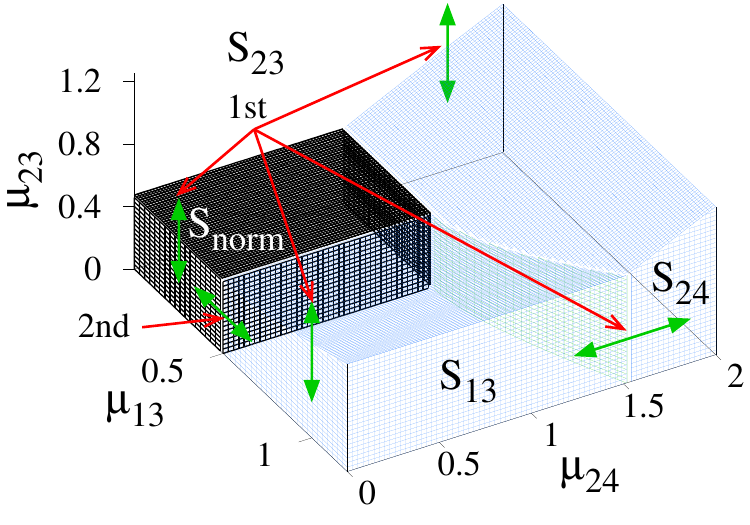}
		\caption{Phase diagram for the {\tt N}-configuration. The normal region is shown in black, with the label $S_{{\rm norm}}$. The collective region is divided by a separatrix (blue surface) below which $\Omega_1$ contributes to the atomic transitions, and above which mode $\Omega_2$ contributes. The region where mode $\Omega_1$ dominates is itself divided by a separatrix (green surface) which determines which of the $2$ subsystems is excited. The parameters used are $\Omega_1=1,\,\Omega_2=0.25,\,\omega_1=0,\, \omega_2=0.8,\,\omega_3= 1$, $\omega_4=1.9$.}
		\label{fsepN}
	\end{center}
\end{figure}

By studying these $2$- and $3$-level subsystems one can reconstruct the phase diagram for any desired configuration. We show that of the $4$-level {\tt N} configuration in Figure~\ref{fsepN}. The normal region is shown in black, with the label $S_{{\rm norm}}$. The collective region is divided by a separatrix (blue surface) below which (labels $S_{13}$ and $S_{24}$) $\Omega_1$ contributes to the atomic transitions, and above which (label $S_{23}$) mode $\Omega_2$ contributes. The region where mode $\Omega_1$ dominates is itself divided by a separatrix (green surface) which determines which of the $2$ subsystems, $S_{13}$ or $S_{24}$, is excited. The transition between the normal region and $S_{13}$ is a second order transition; all others are first order transitions.

The fact that these $2$-level reductions can be carried out iteratively, plus the fact that the polychromatic collective region of the phase space divides itself into monochromatic sub-regions, allows us to overcome the strongest limitation of all: that of the exploding dimension of the Hilbert space when the number of atoms $N$ or the number of excitations $M$ grow. This we treat in the following section.

\section{Reduced Bases}
\label{sec7}

Perhaps the strongest limitation of all, in the study of finite matter-radiation systems, is the fact that the dimension of Hilbert space $\mathcal{H}$ becomes unwieldy as the number of atoms $N$ and/or the number of excitations $M$ grow. Table~\ref{DimensionsTable} shows the dimension of $\mathcal{H}$ for a $3$-level $\Lambda$ configuration under resonant conditions $\Delta_{jk}=0$. The way to read it is as follows:

The first column shows the number of atoms, from $1$ to $5$. The columns labeled ${\rm e}_{10}$ show the dimension required in order for the calculated ground state to differ by less than an error of ${\rm e}^{-10}$ from the exact quantum ground state, as measured by the fidelity between the states. The same goes for the columns labeled ${\rm e}_{15}$, in this case for an error less than ${\rm e}^{-15}$. The numbers in parenthesis at the top of the columns show the value of the dimensionless coupling constant $x_{ij} = \mu_{ij}/\mu^{c, coh}_{ij}$ (where $\mu^{c, coh}_{ij}$ is the critical value of the coupling constant $\mu_{ij}$ using coherent states) at which the dimension is calculated.

It is important to stress that the fidelity constraint is arbitrary, of course, and may be set according to the problem to be tackled; we have chosen these approximations because, for instance, to an error of ${\rm e}^{-10}$ the expectation value of the energy of the ground state remains fixed up to $10^{-8}$ (even for large values of the coupling constants).

\begin{table}
		\caption{Dimension of the Hilbert space $\mathcal{H}$ for a $3$-level $\Lambda$ configuration under resonant conditions $\Delta_{jk}=0$, at different values of the dimensionless coupling constant and for different approximations. See text for details.}
		\label{DimensionsTable}
\begin{center}
\begin{tabular}{c | r | r || r  | r }
$N$ & ${\rm e}_{10}$  $(1.5,1.5)$ & ${\rm e}_{10}$  $(3,3)$ & ${\rm e}_{15}$ $(1.5,1.5)$ &${\rm e}_{15}$ $(3,3)$ \\[1mm] \hline\hline & & &\\[-2mm]
1 & 131 \phantom{aaa}& 397 \phantom{a}&  246 \phantom{aaa}& 584 \phantom{a} \\
2 & 527 \phantom{aaa}& 1\,442 \phantom{a}& 839  \phantom{aaa}& 2\,207 \phantom{a}\\
3 & 1\,058 \phantom{aaa}& 3\,557 \phantom{a}& 1\,622  \phantom{aaa}& 5\,645 \phantom{a}\\
4 & 2\,073 \phantom{aaa}& 7\,797 \phantom{a}&  3\,576 \phantom{aaa}& 12\,552 \phantom{a} \\
5 & 3\,399 \phantom{aaa}& 14\,421 \phantom{a}& 5\,649  \phantom{aaa}& 21\,951 \phantom{a}\\[1mm] \hline
\end{tabular}
\end{center}
\end{table}

The figures differ only slightly for the other configurations $\Xi$ and $V$. This table begs the question, {\it can one reduce the dimension of the Hilbert space while still obtaining essentially the same results as with the exact basis?} The logic behind a possible answer in the affirmative is twofold: 
\begin{itemize}
	\item[i)]
We have iterative method for reducing a system of $n$-level atoms interacting with radiation to a system of $(n-1)$-level atoms. By using repeatedly this method we arrive at a collection of $2$-level subsystems. Thus, looking at the number of atoms to be allowed in each of the $2$-level subsystems is essential.
	\item[ii)]
The polychromatic phase diagram divides itself naturally into monochromatic subregions, where a single electromagnetic mode dominates. Then, checking the total number of excitations allowed in each of the two $2$-level subsystems will be crucial.
\end{itemize}

Once having reduced the system to the study of $2$-level subsystems, each interacting with one mode of the electromagnetic field, the Hamiltonian of each subsystem ${jk}$ possesses only one parity operator
\begin{equation}
\Pi_{jk}=e^{i\pi M_{jk}}\,, \quad M_{jk}=\nu_{jk} + A_{kk}\,,
\label{eq.pi2l}
\end{equation}
with $M_{jk}$ the total number excitations operator for the sub-system ${jk}$ (which would be a constant of motion if the rotating wave approximation were to be considered). From the variational calculation~\cite{nahmad-achar15}, this system presents a phase transition at 
\begin{displaymath}
\bar{\mu}_{jk}^c := \frac{1}{2}\,\sqrt{\Omega_{jk} \,\omega_{kj}}\,; \qquad \omega_{kj}:=\omega_k-\omega_j\, ,
\end{displaymath}
where $j<k$. 

The field basis states are just the corresponding Fock states $\{\vert \nu_{jk} \,\rangle\}$, and if we require the ground state to be unchanged in, say, one part in $10^{-10}$,  a maximum number of photons will be given by a corresponding maximum eigenvalue $m_{jk}$ of $M_{jk}$ that conforms to the desired approximation (which will depend of course on the matter-field coupling strength). For instance, if we are in the $S_{12}$ sector of a $3$-level atom we take $\nu_{12}\leq m_{12}(x_{12})$, and for the other transitions we propose~\cite{cordero19},
\begin{eqnarray}
B_{F_{12}}({\cal O})&=& \{ \vert \nu_{12}\, \nu_{13} \,\nu_{23} \rangle \,  \vert  \ \nu_{12}\leq m_{12}(x_{12}) \, , \nonumber \\
&&\nu_{13}\leq \min\{2 \, {\cal O}+1\, , m_{13}(x_{13})\} \, ,\nonumber\\
&&\nu_{23}\leq \min\{2 \, {\cal O}+1\, , m_{23}(x_{23})\} \} \, .
\end{eqnarray}
with the order ${\cal O}$ in the interval
\begin{equation}
0 \leq {\cal O} \leq\max\left\{ \left\lfloor \frac{m_{12}(x_{12})}{2}\right\rfloor  ,\left\lfloor \frac{m_{13}(x_{13})}{2}\right\rfloor  , \left\lfloor \frac{m_{23}(x_{23})}{2}\right\rfloor\right\} \,.
\label{Ovalue}
\end{equation}
Idem for the other subregions of the collective behaviour. We thereby obtain an ordered sequence of reduced bases for the electromagnetic field, that can be written as the direct sum of the basis states for the different subregions,
\begin{eqnarray}
B_F({\cal O}) &:=& B_{F_{12}}({\cal O}) 
 \oplus \, B_{F_{13}}({\cal O}) \oplus B_{F_{23}}({\cal O}) \,.
\end{eqnarray}
A similar procedure may be followed for the matter sector~\cite{pra}. The complete reduced bases are obtained by their tensorial product with the matter basis, ${\cal B}_\sigma ({\cal O}) = B_F({\cal O}) \otimes B_M({\cal O})$, with ${\cal O}$ indicating the approximation order.

Thus, guided by the ground state variational solution in terms of coherent states, by the constants of motion of the system, and by a fidelity criterion, a sequence of ever-approximating reduced bases may be constructed that has proven to be useful in the study of finite phase diagrams for a finite number of atoms, even when this is large as well as the number of excitations. This allows for the study of previously intractable systems.

\begin{figure}
\begin{center}
\includegraphics[width=0.49\linewidth]{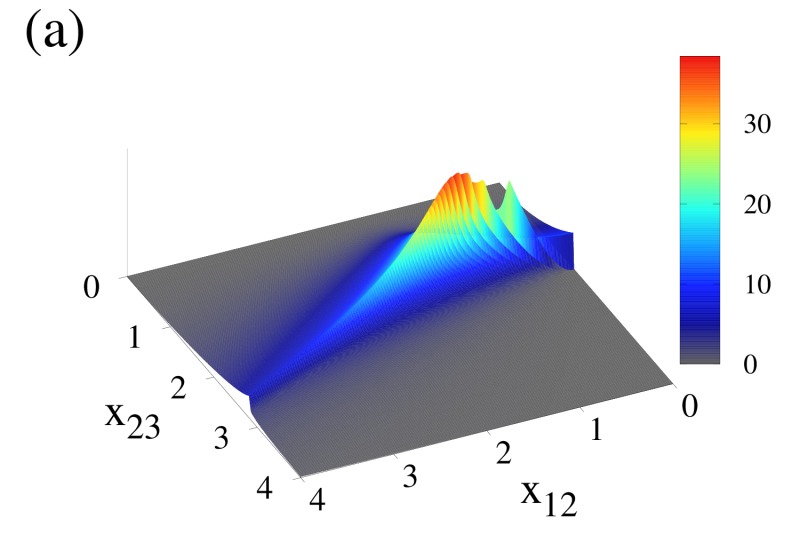}
\includegraphics[width=0.49\linewidth]{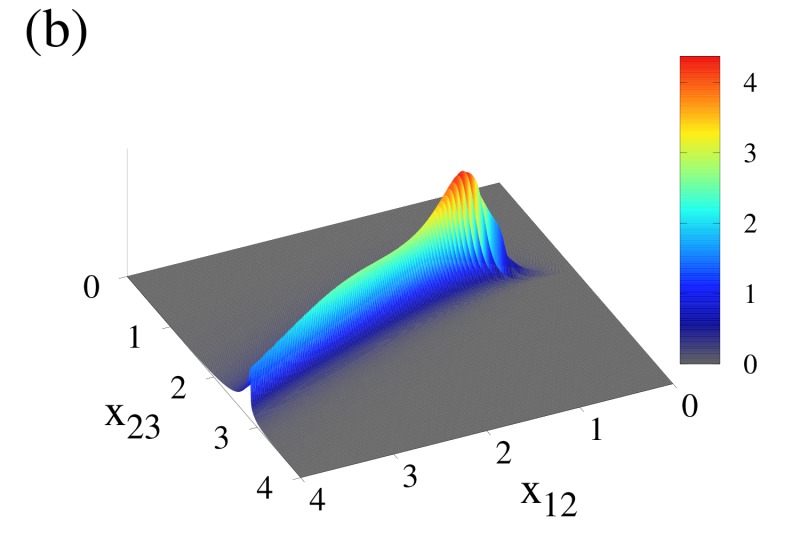}\\
\includegraphics[width=0.49\linewidth]{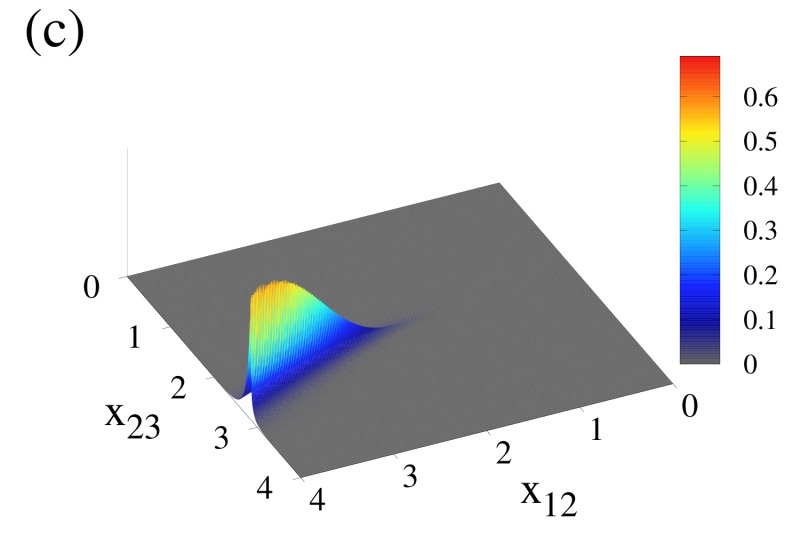}
\end{center}
\caption{Percentual error $\Delta({\cal O})$ in the quantum ground energy surface for the reductions: (a) ${\cal B}_\sigma (0)$, (b)  ${\cal B}_\sigma (1)$ and (c)  ${\cal B}_\sigma (2)$, for a system of $4$ atoms in the $\Xi$ configuration under resonant conditions $\Delta_{jk}=0$. We have set $\omega_1=0,\ \omega_2=1/4,\ \omega_3=1;\ \Omega_1=1/4,\ \Omega_2=3/4$.  Note the different scales in each plot.}
\label{fig8}
\end{figure}

As an example, Table~\ref{tabla3} shows the dimension of the Hilbert space $\mathcal{H}$ for a system of $4$ atoms in the $\Xi$ configuration under resonant conditions $\Delta_{jk}=0$, for the bases of orders $0,\,1,\,2$ and for the exact basis. 
%
\begin{table}
		\caption{Dimension of the Hilbert space $\mathcal{H}$ for a system of $4$ atoms in the $\Xi$ configuration under resonant conditions $\Delta_{jk}=0$, using bases of order $0,\,1,\,2$ and the full (exact) basis.}
		\label{tabla3}
\begin{center}
\begin{tabular}{c | c }
basis & dimension \\ \hline &\\[-3mm]
${\cal B}_\sigma (0)$ & $1,020$  \\ [1mm]
 ${\cal B}_\sigma (1)$ & $2,413$ \\ [1mm]
 ${\cal B}_\sigma (2)$ & $3,609$  \\ [1mm]
${\cal B}_\sigma (exact)$ & $9,546$
\end{tabular}
\end{center}
\end{table}
In Figure~\ref{fig8} the percentual error $\Delta({\cal O})$ in the quantum ground energy surface for each of the reductions is shown, defined as
$$
	\Delta({\cal O}) = \left| \frac{{\cal E}_g - E_{\cal O}}{{\cal E}_g} \right|
$$
where ${\cal E}_g$ denotes the energy of the ground state using the exact basis and $E_{\cal O}$ denotes that obtained from the basis of order ${\cal O}$. We also set $\Delta({\cal O})=0$ when ${\cal E}_g=0$ since all bases give $E=0$ when ${\cal E}_g=0$.
The maximum error is always obtained in a vicinity of the separatrix, as is expected. The difference in scale in each subfigure makes it evident that, as the order increases, ${\cal B}_\sigma ({\cal O})$ is a much better approximation to the exact solution. In fact, for ${\cal B}_\sigma (1)$ we obtain the exact solution in almost all, but not quite, the Normal region. All of the Normal region and much more of the phase space coincides exactly for ${\cal B}_\sigma (2)$, with a maximum error of $0.6\%$ in a very small portion of the separatrix, and yielding a reduction of almost two thirds in the dimension of the Hilbert space.

\section*{Remarks and Conclusions}

We have covered the study of phase diagrams for systems consisting of the interaction of matter with radiation fields, using variational methods based on coherent states that compare very well with the exact quantum solutions, as well as providing analytical expressions for their analysis. We have also shown how restoring the symmetries of the Hamiltonian in the catastrophe formalism improves the agreement with the quantum diagonalisation calculation. In general, $n$-level systems in the presence of $\ell$ electromagnetic modes have been studied. Using the results of the analyses and the behaviour of the solutions we were able to construct a sequence of ever-approximating reduced bases, which make possible the study of larger systems both, in the number of atoms and in the number of excitations. These studies are of importance in fundamental quantum optics, quantum information, and quantum cryptography scenarios.

The work discussed here would not have been possible without the pioneering work of Prof. Roy Glauber using coherent states, who laid the groundwork for the understanding of the behaviour of light from different sources and for new technologies based on quantum optics. He also pioneered the study of first-order phase transitions in statistical physics, and the quantum mechanical behaviour of trapped wave packets. We present this work in his honour.


\section*{Acknowledgements}
We wish to thank Octavio Casta\~nos for useful discussions. This work was partially supported by DGAPA-UNAM under project IN101217.

\end{document}